\documentclass[aps,prc,superscriptaddress,floats,floatfix,reprint,reprintnumbers]{revtex4}
\usepackage{blindtext}
\usepackage{epsfig}
\usepackage{graphicx}% Include figure files
\usepackage{dcolumn}% Align table columns on decimal point
\usepackage{bm}% bold math
\usepackage{booktabs}
\usepackage{color}
\usepackage{mathrsfs}
\usepackage{ulem}
\usepackage{epstopdf}
\graphicspath{{plots/}}%directory of my plots
\usepackage{verbatim}
\usepackage{indentfirst}
\usepackage{float}
\usepackage[colorlinks,
           bookmarks=true,
           linkcolor=blue,
           urlcolor=blue,
            anchorcolor=black,
            citecolor=blue]{hyperref}
\usepackage[bf,raggedright,indentafter]{titlesec}
\usepackage{rotating}
\usepackage{booktabs}
\usepackage{tabularx}
\usepackage{makecell}
\usepackage{hypcap}
\usepackage{subfigure}%
\usepackage{tabularx}

\usepackage{array}

\begin{document}

\hspace{3cm}
\title{\large Studying  the tensor resonance contributions in $B \to PP\ell^+\ell^-$ and $B \to PV\ell^+\ell^-$  decays}

\author{Ru-Min Wang$^{1,\sharp}$,~Xiu-Ping Fan$^{1,\S}$,~Si-Yu Xu$^{1,\natural}$,~Yi  Qiao$^{2,\P}$,~Xiao-Dong Cheng$^{3,\dag}$, and Yuan-Guo Xu$^{1,\ddag}$ \\
{\scriptsize $^1$School of Physics, Jiangxi Normal University, Nanchang, Jiangxi 330022, China}\\
{\scriptsize $^2$College of Physics and Electronic Information, Nanchang Normal University, Nanchang, Jiangxi 330032, China}\\
{\scriptsize $^3$College of Physics and Electronic Engineering, Xinyang Normal University, Xinyang 464000, China}\\
{\scriptsize $^\sharp$ruminwang@sina.com~~~~~~~ $^\S$xiupingfan@yeah.net ~~~~~~$^\natural$siyuxu2025@163.com $^\P$yiqiaophys@163.com~~~~$^\dag$chengxd@mails.ccnu.edu.cn~~~~$^\ddag$yuanguoxu@jxnu.edu.cn}}

\vspace{3cm}

\begin{abstract}
We analyze the semileptonic $B \to T\ell^+\ell^-$,  $B \to T(\to PP)\ell^+\ell^-$, and $B \to T(\to PV)\ell^+\ell^-$ decays  with $\ell=e,\mu,\tau$ based on flavor SU(3) analysis in the standard model ($T$ denotes the light tensor meson, $P$ denotes the light pseudoscalar meson, and $V$ denotes the light vector meson).   The hadronic amplitudes of the $B \to T\ell^+\ell^-$ decays are related by the nonperturbative parameters, and all branching ratios of the $B \to T\ell^+\ell^-$ decays are obtained by the experimental data of the branching ratio of $B^0_s\to f^{\prime}_{2}(1525)\mu^+\mu^-$ in three cases, and then the branching ratios of the  $B \to T(\to PP)\ell^+\ell^-$  and $B \to T(\to PV)\ell^+\ell^-$ decays are predicted by the narrow width approximation and further considering finite width effects of the intermediate resonances.
Compared with the narrow width results,   the finite width effects  slightly reduce the branching fractions for most  decays. However,   sizeable finite width effects are found in some near threshold modes.  For the subthreshold $K_2^*(1430)\to K\eta'$ relevant channels, the finite width of the tensor resonance can open a nonzero contribution.
Compared with the measured $B \to PP\ell^+\ell^-$  and $B \to PV\ell^+\ell^-$ decays, we find that the branching ratios  with the  tensor resonance states are small. Therefore, other resonances, for example, the vector mesons, the scalar mesons, the axial-vector mesons or their excited states, might give the dominant contributions to the relevant decays.
Our results might be tested in current and future experiments.
\end{abstract}
\maketitle

\hspace{2cm}

%\newpage
\section{INTRODUCTION}
Flavor changing neutral current $b\to s/d\ell^+\ell^-$ processes  play a crucial role in testing the standard model and probing new physics beyond it.
In recent years, flavor anomalies have been reported  in  $b\to s\ell^+\ell^-$ quark level transition decays.
Four-body decays $B\to MM\ell^+\ell^-$ with $M=P,V$   are   important as backgrounds to precision analyses in corresponding  benchmark three-body decays, and they can provide a wealth of information on the weak interactions.
Until now,  only  the $B^0_s\to f^{\prime}_{2}(1525)\mu^+\mu^-$ decay has been measured in all $B\to T\ell^+\ell^-$ decays \cite{PDG2025}
\begin{eqnarray}
\mathcal{B}(B^0_s\to f^{\prime}_{2}(1525)\mu^+\mu^-)= (1.62\pm0.22)\times10^{-7}, \label{Eq:fpDATA}
\end{eqnarray}
 some  $B\to PP\ell^+\ell^-$ and $B\to PV\ell^+\ell^-$ rare decays  have been measured by the LHCb Collaboration \cite{LHCb:2014yov}, and the total branching ratios are found to be
 \cite{PDG2025}
\begin{eqnarray}
\mathcal{B}(B^+\to \phi K^+ \mu^+\mu^-)&=(7.9^{+2.1}_{-1.7})\times10^{-8},\nonumber\\
\mathcal{B}(B^0\to \pi^+\pi^- \mu^+\mu^-)&=(2.1\pm0.5)\times10^{-8},\nonumber\\
\mathcal{B}(B^0_s\to \pi^+\pi^- \mu^+\mu^-)&=(8.4\pm1.7)\times10^{-8}.\label{eq:4BodyE}
\end{eqnarray}
Moreover, the S-wave, P-wave and  D-wave contributions to $B^{0}\rightarrow K^{+}\pi^{-}\mu^{+}\mu^{-}$  have been measured in the $644 < m(K^+\pi)<1200 \mbox{MeV}/c^2$, $796 < m(K^+\pi)<996 \mbox{MeV}/c^2$, and $1330 < m(K^+\pi)<1530 \mbox{MeV}/c^2$ regions \cite{LHCb:2016ykl,LHCb:2020lmf,LHCb:2015svh,LHCb:2013zuf,LHCb:2016eyu}. Based on these experimental results, we can conduct preliminary studies on the relevant decay processes.

The $B\to T$ transition  form factors   have been evaluated by using many approaches, such as  ISGW model \cite{Isgur:1988gb}, ISGW2 model \cite{Scora:1995ty,Sharma:2010yx}, the perturbative QCD (PQCD) \cite{Wang:2010ni}, the light cone sum rule (LCSR) \cite{Aliev:2024rea,Yang:2010qd,Aliev:2019ojc,Zuo:2021kui}, the QCD sum rule \cite{Khosravi:2015jfa}, and the large energy effective theory  approach \cite{Charles:1998dr}.
As for the $B\to MM$  form factors,  it is rather challenging to study in  both experimental and theoretical contexts. Until now,  only S-wave $B\to K\pi$ form factors have been obtained with Light-Cone Sum Rules  \cite{Descotes-Genon:2023ukb},   and the lack of reliable $B\to MM$  form factors means that the physics potential of this component
remains untapped.  In the absence of reliable calculations,  symmetry analysis plays a vital  role  by providing relations between hadronic form factors or hadronic decay amplitudes.
 SU(3) flavor symmetry is one of the symmetries which has
attracted a lot of attention. It has been widely used to study hadron decays, for instance,  b-hadron decays \cite{Qiao:2025rhm,Wang:2021uzi,He:1998rq,He:2000ys,Fu:2003fy,Hsiao:2015iiu,He:2015fwa,He:2015fsa,Deshpande:1994ii,Gronau:1994rj,Gronau:1995hm,Shivashankara:2015cta,Zhou:2016jkv,Cheng:2014rfa,Wang:2020wxn}.

The $B\to T\ell^+\ell^-$ decays  such as  $B \to K^*_2(1430)\ell^+\ell^-$ has been extensively analyzed \cite{Das:2018orb,Aliev:2011gc,Katirci:2011mt,Junaid:2011egj,Li:2010ra,Choudhury:2009fz,Ahmed:2012zzc,Zuo:2021kui,Vardani:2024bae}.
Some $B\to T(T\to PP,PV)\ell^+\ell^-$ decays  have been studied in the standard model and some new physics models, for example, $B\to K^*_J (\to K \pi) \mu^+\mu^-$  decay  \cite{Lu:2011jm,Gratrex:2015hna,Kim:2007fx,Capdevila:2025drq}, and $B_s\to f_{2}'(1525)(\to K^+\,K^-)\mu^+\mu^-$ decays  \cite{Rajeev:2020aut}.   In addition,
P- and S-wave contributions to the $B^{0}\to K^{+}\pi^-\ell^+\ell^-$  have been studied in Ref. \cite{Alguero:2021yus}, and the model-independent distributions for non-resonance $ \overline{B}\to \overline{K}\pi \ell \ell $ and $ {\overline{B}}_s\ \to \overline{K}K\ell\ell$  can be found in Ref. \cite{Das:2014sra}.

In this work, we conduct a comprehensive branching ratio analysis of the $B\to T(\to PP)\ell^+\ell^-$ and $B\to T(\to PV)\ell^+\ell^-$ decays by using the SU(3) flavor symmetry approach and assisting  with the $B\to T$ form factors  from PQCD and LCSR.
The analysis begins by obtaining the form factor relations for the  $B \to T \ell^+ \ell^-$ decays through  SU(3) flavor symmetry/breaking.  Owing to the lack of relevant experimental data, the branching ratios of the $B \to T\ell^+ \ell^-$ decays are predicted by only considering  the SU(3) flavor symmetry contribution.
Then using  these estimates  along with  previous results for $T\to PP,PV$  decays, the branching ratios for $B\to T(T\to PP)\ell^+\ell^-$ and $B\to T(T\to PV)\ell^+\ell^-$  are obtained by using  the narrow width approximation and  further considering finite width effects of the intermediate resonances.

The paper is structured  as follows.  In Section II, the three-body semileptonic charmless  $B\to T\ell^+\ell^-$ decays are studied.  In Section III, the four body semileptonic charmless  $B\to PP\ell^+\ell^-,PV\ell^+\ell^-$ with   the tensor  resonance states  are explored. Summary and conclusions are presented in Section IV.

%\newpage
\section{Semileptonic charmless  $B\to T\ell^+\ell^-$ decays}\label{sec:B2Mlv}

\subsection{Theoretical framework}

In the standard model, the low energy effective Hamiltonian for  $b\to q'\ell^+\ell^- (q'=s,d)$ transitions  can be written as  \cite{Li:2010ra}
\begin{eqnarray}
\mathcal{H}(b\to
 q'\ell^+\ell^-)&=&\frac{G_F}{\sqrt2}\frac{\alpha_{\rm e}}{\pi}V_{tb}V_{tq'}^*\times
 \left\{ \frac{C_9+C_{10}}{4}[\bar q'b]_{V-A}[\bar \ell \ell]_{V+A}
 +\frac{C_9-C_{10}}{4}[\bar q'b]_{V-A}[\bar \ell \ell]_{V-A}\right. \nonumber\\
 &&\left.~~~~~~~~~~~~~~~~~~+ C_{7L}m_b[\bar q' i\sigma_{\mu\nu}
 (1+\gamma_5)b]\frac{q^\mu}{q^2}\times[\bar \ell \gamma^\nu \ell]+ C_{7R}m_b[\bar q' i\sigma_{\mu\nu}
 (1-\gamma_5)b]\frac{q^\mu}{q^2}\times[\bar \ell \gamma^\nu \ell]\right\},\label{eq:H}
\end{eqnarray}
where $G_F$ denotes the Fermi constant, $\alpha_e=\frac{e^2}{4\pi}$ is the fine structure constant, $(\overline{f}_1f_2)_{V\pm A}=\overline{f}_1\gamma_\mu(1\pm\gamma_5)f_2$,  $\sigma_{\mu\nu}=\frac{i[\gamma_\mu,\gamma_\nu]}{2}$, and $V_{tb}V^*_{tq'}$  denotes the Cabibbo-Kobayashi-Maskawa (CKM) elements.  Wilson coefficients $C_{7L}=C_7$ and $C_{7R}=\frac{m_{q'}}{m_b}C_{7}$, and  the expressions of $C_{7,9}$ as well as $C_{10}$  are taken from  Ref. \cite{Buchalla:1995vs}.

In analogy with $B\to V$ form factors, the hadronic matrix elements  for  the $B\to T$   transition are parameterized in terms of seven form factors $V(q^2)$, $A_{0,1,2}(q^2)$  and $T_{1,2,3}(q^2)$ \cite{Yang:2010qd,Hatanaka:2009gb,Hatanaka:2010fpr,Wang:2010ni}
 \begin{eqnarray}
\langle T(P_2,\epsilon)|\bar q'\gamma^{\mu}b|\overline B(P_B)\rangle                     &=&-\frac{2V(q^2)}{m_B+m_{T}}\epsilon^{\mu\nu\rho\sigma} \epsilon^*_{T\nu}  P_{B\rho}P_{2\sigma}, \nonumber\\\langle  T(P_2,\epsilon)|\bar q'\gamma^{\mu}\gamma_5 b|\overline  B(P_B)\rangle   &=&2im_{T} A_0(q^2)\frac{\epsilon^*_{T } \cdot  q }{ q^2}q^{\mu}    +i(m_B+m_{T})A_1(q^2)\left[ \epsilon^*_{T\mu }    -\frac{\epsilon^*_{T } \cdot  q }{q^2}q^{\mu} \right] \nonumber\\
                                                                                         & &-iA_2(q^2)\frac{\epsilon^*_{T} \cdot  q }{  m_B+m_{T} }    \left[ P^{\mu}-\frac{m_B^2-m_{T}^2}{q^2}q^{\mu} \right],\nonumber\\
\langle  T(P_2,\epsilon)|\bar q'\sigma^{\mu\nu}q_{\nu}b|\overline  B(P_B)\rangle         &=&-2iT_1(q^2)\epsilon^{\mu\nu\rho\sigma} \epsilon^*_{T\nu} P_{B\rho}P_{2\sigma}, \nonumber\\
\langle  T(P_2,\epsilon)|\bar q'\sigma^{\mu\nu}\gamma_5q_{\nu}b|\overline  B(P_B)\rangle &=&T_2(q^2)\left[(m_B^2-m_{T}^2) \epsilon^*_{T\mu }   - {\epsilon^*_{T } \cdot  q }  P^{\mu} \right] +T_3(q^2) {\epsilon^*_{T } \cdot  q }\left[ q^{\mu}-\frac{q^2}{m_B^2-m_{T}^2}P^{\mu}\right],\label{Eq:FF}
 \end{eqnarray}
where $q=P_B-P_2, P=P_B+P_2$, and $\epsilon$ denotes  the polarization vector of the tensor mesons.

The hadronic amplitude for the $B\to T \ell^+\ell^-$ processes can be obtained from the  effective Hamiltonian in Eq. (\ref{eq:H}) and  the form factors in Eq. (\ref{Eq:FF}) \cite{Li:2010ra}
\begin{eqnarray}
 A_{L0}  &=& N_{T}  \frac{\sqrt{\lambda}}{\sqrt6 m_Bm_{T}}\frac{1}{2m_{T}\sqrt {q^2}}\left[ (C_9-C_{10})[(m_B^2-m_{T}^2-q^2)(m_B+m_{T})A_1 -\frac{\lambda}{m_B+m_{T}}A_2]\right.\nonumber\\
           &&\left. +  2m_b(C_{7L}-C_{7R})  [ (m_B^2+3m_{T}^2-q^2)T_2 -\frac{\lambda  } {m_B^2-m_{T}^2}T_3]\right],\nonumber\\
A_{L\perp}&=& -\sqrt{2} \frac{\sqrt{\lambda}}{\sqrt8m_Bm_{T}}N_{T}\left[(C_9-C_{10}) \frac{\sqrt \lambda V}{m_B+m_{T}}+\frac{2m_b(C_{7L}+C_{7R})}{q^2}\sqrt \lambda T_1\right],\nonumber\\
A_{L||}&=& \sqrt{2}\frac{\sqrt{\lambda}}{\sqrt 8m_Bm_{T}} N_{T} \left[(C_9-C_{10}) (m_B+m_{T})A_1+\frac{2m_b(C_{7L}-C_{7R})}{q^2}(m_B^2-m_{T}^2) T_2 \right],\nonumber\\
 A_{Lt}&=& N_{T}   \frac{\sqrt{\lambda}}{\sqrt 6m_Bm_{T}} (C_{9}-C_{10})\frac{\sqrt \lambda}{\sqrt {q^2}}A_0,\nonumber\\
A_{Ri}  &=& A_{Li}|_{C_{10}\to -C_{10}}, \nonumber\\
 A_t&=&A_{Rt}-A_{Lt}= 2N_{T} \frac{\sqrt{\lambda}}{\sqrt6m_Bm_{T}}C_{10}\frac{\sqrt \lambda}{\sqrt {q^2}}A_0.\label{Eq:HA}
\end{eqnarray}
with  $N_{T}=\Big[\frac{\tau_BG_F^2 \alpha_{\rm e}^2}{3\cdot 2^{10}\pi^5
m_B^3}|V_{tb}V_{tq'}^*|^2 q^2\lambda^{1/2}
\left(1-\frac{4m_\ell^2}{q^2}\right)^{1/2}\Big]^{1/2}$ \cite{Mohapatra:2021izl}, and  please note that we put $\tau_B$ in $N_{T}$.

Various interesting observables for the processes $B\to T\ell^+\ell^-$   are given as follows \cite{Li:2010ra,Descotes-Genon:2012isb,Descotes-Genon:2013vna}\\
$\bullet$ Differential decay branching ratio:
\begin{eqnarray}
 \frac{ d\mathcal{B}}{dq^2} = \frac{1}{4} \left(3I_1^c+6I_1^s-I_2^c-2I_2^s\right), \label{Eq:dBr}
\end{eqnarray}
$\bullet$ Differential  longitudinal polarization fraction:
\begin{eqnarray}
 f_L(q^2)= \frac{3I_1^c-I_2^c}{3I_1^c+6I_1^s-I_2^c-2I_2^s},
\end{eqnarray}
$\bullet$ Differential forward-backward asymmetry of lepton pair:
\begin{eqnarray}
 A_{FB}(q^2)= \frac{3I_6}{3I_1^c+6I_1^s-I_2^c-2I_2^s},
\end{eqnarray}
$\bullet$ Some other observables with reduced uncertainty
\begin{eqnarray}
<P_1>&=&\frac{1}{2} \frac{\int dq^2I_3}{\int dq^2I_2^s},~~~~~~~~~~~~~~~~~~~~<P_2>=\frac{1}{8} \frac{\int dq^2I_6^s}{\int dq^2I_2^s},\nonumber\\
<P'_4>&=&\frac{\int dq^2I_4}{\sqrt{-\int dq^2I_2^s \int dq^2I_2^c}},~~~~~~<P'_5>=\frac{\int dq^2I_5}{2\sqrt{-\int dq^2I_2^s \int dq^2I_2^c}},
\end{eqnarray}
where the angular coefficients are
\begin{eqnarray}
 I_1^c&=&  (|A_{L0}|^2+|A_{R0}|^2) +8 \frac{m_\ell^2}{q^2}{\rm Re}[A_{L0}A^*_{R0} ]+4\frac{m_\ell^2}{q^2} |A_t|^2, \nonumber\\
 I_1^s&=&\frac{3}{4} [|A_{L\perp}|^2+|A_{L||}|^2+|A_{R\perp}|^2+|A_{R||}|^2  ] \left(1-\frac{4m_\ell^2}{3q^2}\right)+\frac{4m_\ell^2}{q^2} {\rm Re}[A_{L\perp}A_{R\perp}^* + A_{L||}A_{R||}^*],\nonumber\\
 I_2^c&=& -\beta_\ell^2(  |A_{L0}|^2+ |A_{R0}|^2),\nonumber\\
 I_2^s&=& \frac{1}{4}\beta_\ell^2(|A_{L\perp}|^2+|A_{L||}|^2+|A_{R\perp}|^2+|A_{R||}|^2), \nonumber\\
 I_3  &=&\frac{1}{2}\beta_\ell^2(|A_{L\perp}|^2-|A_{L||}|^2+|A_{R\perp}|^2-|A_{R||}|^2),\nonumber\\
 I_4  &=& \frac{1}{\sqrt2}\beta_\ell^2  [{\rm Re}(A_{L0}A_{L||}^*)+{\rm  Re}(A_{R0}A_{R||}^*)],~~~~~~~~~~~~~~~~~~~~I_5  = \sqrt 2\beta_\ell  [{\rm Re}(A_{L0}A_{L\perp}^*)-{\rm Re}(A_{R0}A_{R\perp}^*)],\nonumber\\
 I_6  &=& 2\beta_\ell  [{\rm Re}(A_{L||}A^*_{L\perp})-{\rm  Re}(A_{R||}A^*_{R\perp})],~~~~~~~~~~~~~~~~~~~~~ I_7 = \sqrt2\beta_\ell  [{\rm Im}(A_{L0}A^*_{L||})-{\rm Im}(A_{R0}A^*_{R||})],\nonumber\\
 I_8  &=& \frac{1}{\sqrt2}\beta_\ell^2  [{\rm Im}(A_{L0}A^*_{L\perp})+{\rm  Im}(A_{R0}A^*_{R\perp})],~~~~~~~~~~~~~~~~~~ I_9 =\beta_\ell^2  [{\rm Im}(A_{L||}A^*_{L\perp})+{\rm  Im}(A_{R||}A^*_{R\perp})], \label{eq:AF}
\end{eqnarray}
with $\beta_\ell=\sqrt{1-4m^2_\ell/q^2}$. For the later numerical results of $f_L$ and  $A_{FB}$,
the normalized $q^2$ integrated are obtained  by separately integrating the numerators and denominators with the same $q^2$ bins.

Then the observables of the $B\to T\ell^+\ell^-$ decays could be obtained by the relevant form
factors, which depend on different methods.
The SU(3) flavor symmetry approach is independent of
the detailed dynamics, and it can provide the hadronic
amplitude  relationships among different decay modes, and their form factors also follow the same relationships. In this work, we will use SU(3) flavor symmetry to analyze these
decays.

\subsection{SU(3) hadronic amplitude relations }
 The SU(3) flavor analysis depends on
the SU(3) flavor group, and the relevant meson multiplets are listed as follows.
Bottom pseudoscalar triplet $B_i$ is
\begin{eqnarray}
B_i=\Big(B^+(\bar{b}u),B^0(\bar{b}d),B^0_s(\bar{b}s)\Big).
\end{eqnarray}
The octet and the singlet  of p-wave tensor mesons with $J^{PC}=2^{++}$ take the form   \cite{Ecker:2007us,Chen:2023ybr}
\begin{eqnarray}
 T&=&\left(\begin{array}{ccc}
\frac{a_2^0}{\sqrt{2}}+\frac{f_2^q}{\sqrt{2}} & a^+_2 & K_2^{*+}  \\
a^-_2 & -\frac{a^0_2}{\sqrt{2}}+\frac{f^q_2}{\sqrt{2}} & K^{*0}_2 \\
K^{*-}_2 & \overline{K}^{*0}_2 & f^s_2
\end{array}\right)\,.
\end{eqnarray}
where the tensor mesons are  composed of quark-antiquark  pairs.
In this paper, the isovector mesons $a_2(1320)$, isodoublet states $K_2^*(1430)$ and two isosinglet mesons $f_2(1270)$ and  $f_2'(1525)$ are investigated; $f_{2}(1270)$ and $f_2'(1525)$ are mixed by $f_2^q$ and $f_2^s$ with the mixing angle $\theta_{f_2}$, as follows:
\begin{eqnarray}
\left(\begin{array}{c}
f_2(1270)\\
f_2'(1525)
\end{array}\right)\,
=
\left(\begin{array}{cc}
cos\theta_{f_2}&sin\theta_{f_2}\\
sin\theta_{f_2}&-cos\theta_{f_2}
\end{array}\right)\,
\left(\begin{array}{c}
f^q_2\\
f^s_2
\end{array}\right),
\end{eqnarray}
where $~\theta_{f_2}\in[8^\circ, 10^\circ]$ \cite{Cheng:2010yd} will be used in our calculation.

Light pseudoscalar mesons $P$ and vector mesons  $V$ will be used in  Sec. \ref{sec:D4MM}, and they   are given here together \cite{He:2018joe}
\begin{eqnarray}
 P&=&\left(\begin{array}{ccc}
\frac{\pi^0}{\sqrt{2}}+\frac{\eta_8}{\sqrt{6}}+\frac{\eta_1}{\sqrt{3}} & \pi^+ & K^+ \\
\pi^- &-\frac{\pi^0}{\sqrt{2}}+\frac{\eta_8}{\sqrt{6}}+\frac{\eta_1}{\sqrt{3}}  & K^0 \\
K^- & \overline{K}^0 &-\frac{2\eta_8}{\sqrt{6}}+\frac{\eta_1}{\sqrt{3}}
\end{array}\right)\,,\\
%\end{eqnarray}
%\begin{eqnarray}
V&=&\left(\begin{array}{ccc}
\frac{\rho^0}{\sqrt{2}}+\frac{\omega}{\sqrt{2}} & \rho^+ & K^{*+} \\
\rho^- &-\frac{\rho^0}{\sqrt{2}}+\frac{\omega}{\sqrt{2}} & K^{*0} \\
K^{*-} & \overline{K}^{*0} &\phi
\end{array}\right)\,,\label{Eq:VM}
\end{eqnarray}
and the pseudoscalar mesons $\eta$ and $\eta'$can be obtained by
\begin{eqnarray}
\left(\begin{array}{c}
\eta\\
\eta'
\end{array}\right)\,
=
\left(\begin{array}{cc}
\mbox{cos}\theta_P&-\mbox{sin}\theta_P\\
\mbox{sin}\theta_P&\mbox{cos}\theta_P
\end{array}\right)\,\left(\begin{array}{c}
\eta_8\\
\eta_1
\end{array}\right)\,,
\end{eqnarray}
with $\theta_P=[-20^\circ,-10^\circ]$ from the PDG \cite{PDG2025}.

In the $B\to T\ell^+\ell^-$ decays, the leptonic current is invariant under the
SU(3) flavor symmetry, and the hadronic amplitudes can be parameterized by
the SU(3) flavor symmetry/breaking  as
\begin{eqnarray}
A(B\to T\ell^+\ell^-)=a_0B^iM_i^jH_j+a_1B^aW_a^iM_i^jH_j+a_2B^iM_i^aW_a^jH_j+b_0B^iM_j^jH_i,\label{eq:ASU3}
\end{eqnarray}
where $a_0$ and $b_0$ are the nonperturbative coefficients under the
SU(3) flavor symmetry,  the $b_0$ term  corresponds to the annihilation process, and it is  suppressed by the
Okubo-Zweig-Iizuka rule. $a_{1,2}$ are the nonperturbative SU(3) flavor breaking coefficients, and they are also  suppressed. The matrix $W$ is related to
the SU(3) flavor breaking effects due to  different masses of $u,d$ and $s$ quarks \cite{He:2014xha}.
In addition,  $H_1=0$, $H_2=V_{tb}V_{td}^*$ for $b\to d\ell^+\ell^-$ transitions as well as $H_3=V_{tb}V_{ts}^*$ for $b\to s\ell^+\ell^-$ transitions.
It should be emphasized that Eq. (\ref{eq:ASU3}) represents the reduced SU(3) flavor amplitudes. The SU(3) flavor symmetry constrains the flavor structure of the hadronic matrix elements, but it does not determine the Lorentz decomposition or the $q^2$ dependence of the form factors. In the exact SU(3) limit, the same flavor coefficients relate each form factor separately, while the relative magnitudes and $q^2$ dependences of $V$, $A_i$, and $T_i$ have to be supplied by external dynamical inputs.

\begin{table}[ht]
\caption{The hadronic amplitudes for $B \to T\ell^+\ell^-$ decays due to the $b\to d/s\ell^+\ell^-$ transitions under the SU(3) flavor symmetry. }	
\begin{center}    		\renewcommand\arraystretch{1.5}\tabcolsep 0.35in
\begin{tabular}{lc}\hline\hline   			
  Decay modes                                             &  SU(3) hadronic amplitudes                                                         \\ \hline
$B^+_u\to K^{*}_{2}(1430)^+\ell^+\ell^-$                  & $\big[a_0+a_1-2a_2\big]~V_{tb}V_{ts}^*$                                                                       \\
$ B^0_d\to K^{*}_{2}(1430)^0\ell^+\ell^-$                 & $\big[a_0+a_1-2a_2\big]~V_{tb}V_{ts}^*$                                                                       \\
$B^0_s\to f_{2}(1270)\ell^+\ell^-$                        & $\big[sin\theta _{f2}~(a_0-2a_1-2a_2)+({\sqrt{2}}cos{\theta _{f2}}+sin{\theta _{f2}})~b_0\big]~V_{tb}V_{ts}^*$   \\
$B^0_s\to f^{\prime}_{2}(1525)\ell^+\ell^-$               & $\big[-cos\theta _{f2}~(a_0-2a_1-2a_2)+({\sqrt{2}}sin{\theta _{f2}}-cos{\theta _{f2}})~b_0\big]~V_{tb}V_{ts}^*$   \\\hline
$B^+_u\to a_{2}(1320)^+\ell^+\ell^-$                      & $\big[a_0+a_1+a_2\big]~V_{tb}V_{td}^*$                                                                        \\
$B^0_d\to a_{2}(1320)^{0}\ell^+\ell^-$                    & $-\frac{1}{\sqrt{2}}\big[a_0+a_1+a_2\big]~V_{tb}V_{td}^*$                                                    \\
$B^0_d\to f_{2}(1270)\ell^+\ell^-$                        & $\big[\frac{1}{\sqrt{2}}cos\theta _{f2}~(a_0-2a_1+a_2)+(\sqrt{2}cos\theta _{f2}+sin\theta _{f2})~b_0\big]~V_{tb}V_{td}^*$          \\
$B^0_d\to f^{\prime}_{2}(1525)\ell^+\ell^-$               & $\big[\frac{1}{\sqrt{2}}sin\theta _{f2}~(a_0-2a_1+a_2)+(\sqrt{2}sin\theta _{f2}-cos\theta _{f2})~b_0\big]~V_{tb}V_{td}^*$           \\
$B^0_s\to \overline{K}^{*}_{2}(1430)^0\ell^+\ell^-$       & $\big[a_0-2a_1+a_2\big]~V_{tb}V_{td}^*$                                                                         \\ \hline
\end{tabular}
\end{center}\label{Tab:ARB2T}
\end{table}

The hadronic amplitude relations for the $B\to T\ell^+\ell^-$ decays are listed in Tab. \ref{Tab:ARB2T},
and the CKM  element products are also listed in Tab. \ref{Tab:ARB2T} for convenience. The same SU(3) flavor relations hold separately for each form factor in the SU(3) limit.
If neglecting the SU(3) flavor breaking $a_{1,2}$ terms and the OZI suppressed $b_0$ term, all decay modes can be related by only one parameter $a_0$.
So far there are not  many experimental results in the $B\to T\ell^+\ell^-$ decays, and only $\mathcal{B}(B^0_s\to f^{\prime}_{2}(1525)\mu^+\mu^-)$ decay has been measured as given in Eq. (\ref{Eq:fpDATA}).
Since there are not enough experimental data in the $B\to T\ell^+\ell^-$ decays, we therefore consider only the nonperturbative coefficient $a_0$ for the numerical results at present.

It should be stressed that the numerical analysis below is performed in the SU(3) symmetric limit. Although the SU(3) breaking structures  related to  $a_1$ and $a_2$ are shown in Eq.  (\ref{eq:ASU3}), the present experimental information on $B\to T\ell^+\ell^-$ decays is not sufficient to determine these additional  nonperturbative coefficients. Introducing SU(3) breaking effects at the level of $20\%-30\%$ without independent experimental constraints would lead to additional model dependent parameters and substantially enlarge the uncertainties of the branching ratio estimates. Therefore, in the present work we do not attempt a quantitative fit of SU(3) breaking contributions. The numerical results should be regarded as estimates in the SU(3) symmetric limit, while the possible SU(3) breaking corrections represent an additional source of  uncertainty.

\subsection{Numerical results}
The theoretical input parameters, such as the lifetimes, the masses, and the
experimental data within the $1\sigma$ error bar from PDG \cite{PDG2025}
will be used in our numerical analysis.
The numerical analysis is performed in three different schemes, denoted as $S_1$, $S_{\rm PQCD}$, and $S_{\rm LCSR}$. These schemes are designed to separate the effects of SU(3) flavor symmetry from the additional model dependence associated with the $B\to T$ transition form factors. In all three schemes, the available experimental measurement of $\mathcal{B}(B_s^0\to f_2'(1525)\mu^+\mu^-)$ is used as the normalization input. Therefore, the differences among the three schemes mainly reflect the treatment of the hadronic form factors and their $q^2$ dependence.

\subsubsection{Results in $S_1$ scheme}
All hadronic amplitudes are proportional to $N_T \sqrt{\lambda}/(m_Bm_T)$ as given in Eq. (\ref{Eq:HA}), so the
differential decay branching ratio in Eq. (\ref{Eq:dBr}) can be written as
\begin{eqnarray}
 \frac{ d\mathcal{B}}{dq^2} =\frac{N_T^2 \lambda}{4m^2_Bm^2_T}  H_{B\to T}^2,
\end{eqnarray}
with $H_{B\to T}^2=3I'^c_1+6I'^s_1-I'^c_2-2I'^s_2$ and $I'^{q}_i\equiv \frac{m^2_Bm^2_T}{N_T^2 \lambda}I^{q}_i$.
After ignoring  the $m^2_\ell$ terms (this neglect will cause that the results with $\ell=\tau$  are not very accurate, so we do not give the results of the $B\to T\tau^+\tau^-$ decays), $H_{B\to T}$ only includes the hadronic part, and follows the same relationships in Tab. \ref{Tab:ARB2T}.

 In the $S_1$ scheme, we use the SU(3) flavor relations directly at the level of the hadronic amplitudes. After neglecting the terms proportional to $m_\ell^2$, the differential branching ratio can be expressed in terms of an effective hadronic quantity $H_{B\to T}$, which is assumed to be independent of $q^2$. This approximation provides a relatively model independent estimate of the branching ratios for the electron and muon modes. However, since the lepton mass effects are neglected, this scheme is not suitable for reliable estimates of the $\tau^+\tau^-$ modes. The $S_1$ scheme therefore serves as a baseline SU(3) symmetry estimate with minimal dependence on specific form factor calculations.  After considering the bounds of  $\mathcal{B}(B^0_s\to f^{\prime}_{2}(1525)\mu^+\mu^-)$ given in Eq. (\ref{Eq:fpDATA}), $H_{B\to T}$ can be  determined, and we obtain
$H_{B\to T}=24.70\pm2.34$. The estimates of the branching ratios  are obtained and listed in the second column of  Tab. \ref{Tab:data1BrB2TllPred}.

\begin{table}[t]
\caption{The branching ratios for the $B \to T\ell^+\ell^-$ decays within  1$\sigma$ error.  They are in unit of $10^{-7}$ when $\ell=e,\mu$ and   in unit of  $10^{-10}$  when $\ell=\tau$. Previous estimates are obtained by using  the LCSR \cite{Aliev:2019ojc,Junaid:2011egj}, the PQCD \cite{Li:2010ra}, and the LCSR in HQEFT \cite{Zuo:2021kui}.  }
\begin{center}\renewcommand\arraystretch{1.2}
{\footnotesize
\begin{tabular}{lccccc} \hline\hline
                                                                     &          $S_1$           &       $S_{PQCD}$       &        $S_{LCSR}$      &     Previous ones \\ \hline
{\color{blue}$\overline{b}\to \overline{s}\ell^+\ell^-:$}\\
$\mathcal{B}(B^+_u\to K^{*}_{2}(1430)^+e^+e^-)$                      &     $2.03\pm 0.32$       &     $2.90\pm 1.27$     &     $2.37\pm 0.64$     &     $7.72\pm4.28$ \cite{Aliev:2019ojc},  $4.21^{+1.01}_{-0.82}$ \cite{Zuo:2021kui}\\
$\mathcal{B}(B^0_d\to K^{*}_{2}(1430)^0e^+e^-)$                      &     $1.86\pm 0.29$       &     $2.68\pm 1.16$     &     $2.18\pm 0.60$     &     $7.72\pm4.28$ \cite{Aliev:2019ojc},  $3.90^{+0.94}_{-0.75}$ \cite{Zuo:2021kui} \\
$\mathcal{B}(B^0_s\to f_{2}(1270)e^+e^-)$                            &     $0.079\pm 0.028$     &     $0.13\pm 0.06$     &     $0.11\pm 0.04$     &      $...$ \\
$\mathcal{B}(B^0_s\to f^{\prime}_{2}(1525)e^+e^-)$                   &     $1.63\pm 0.22$       &     $2.62\pm 0.83$     &     $2.03\pm 0.41$     &      $3.82^{+1.14}_{-0.81}$ \cite{Zuo:2021kui} \\
$\mathcal{B}(B^+_u\to K^{*}_{2}(1430)^+\mu^+\mu^-)$                  &     $2.01\pm 0.32$       &     $1.97\pm 0.34$     &     $2.01\pm 0.35$     &     $6.05\pm3.81$ \cite{Aliev:2019ojc},  $2.5^{+1.6}_{-1.1}$ \cite{Li:2010ra},\\
&&&& $2.45^{+0.59}_{-0.49}$ \cite{Zuo:2021kui}, $2.43^{+0.6}_{-0.5}$ \cite{Junaid:2011egj}\\
$\mathcal{B}(B^0_d\to K^{*}_{2}(1430)^0\mu^+\mu^-)$                  &     $1.85\pm 0.29$       &     $1.80\pm 0.30$     &     $1.83\pm 0.31$     &     $6.05\pm3.81$ \cite{Aliev:2019ojc}, $2.5^{+1.6}_{-1.1}$ \cite{Li:2010ra}, $2.27^{+0.55}_{-0.44}$ \cite{Zuo:2021kui}, \\
&&&&$2.43^{+0.6}_{-0.5}$ \cite{Junaid:2011egj}, $3.5^{+1.30}_{-1.17}$ \cite{Hatanaka:2009gb} \\
$\mathcal{B}(B^0_s\to f_{2}(1270)\mu^+\mu^-)$                        &     $0.079\pm 0.028$     &     $0.096\pm 0.033$     &     $0.096\pm 0.033$     &     $...$ \\
$\mathcal{B}(B^0_s\to f^{\prime}_{2}(1525)\mu^+\mu^-)$               &     $1.62\pm 0.22$       &     $1.62\pm 0.22$     &     $1.62\pm 0.22$     &     $1.8^{+1.1}_{-0.7}$ \cite{Li:2010ra}, $2.31^{+0.69}_{-0.50}$ \cite{Zuo:2021kui}\\
$\mathcal{B}(B^+_u\to K^{*}_{2}(1430)^+\tau^+\tau^-)$                &     $\cdots$       &     $8.08\pm 2.693$     &     $3.81\pm 1.23$     &      $11.2\pm5.9$ \cite{Aliev:2019ojc},  $9.6^{+6.2}_{-4.5}$ \cite{Li:2010ra},\\
&&&& $1.56^{+0.38}_{-0.32}$ \cite{Zuo:2021kui}, $2.74\pm0.9$ \cite{Junaid:2011egj} \\
$\mathcal{B}(B^0_d\to K^{*}_{2}(1430)^0\tau^+\tau^-)$                &     $\cdots$       &     $7.12\pm 2.56$     &     $3.24\pm 1.01$     &     $11.2\pm5.9$ \cite{Aliev:2019ojc}, $9.6^{+6.2}_{-4.5}$ \cite{Li:2010ra}, \\
&&&&$1.45^{+0.36}_{-0.29}$ \cite{Zuo:2021kui}, $2.74\pm0.9$ \cite{Junaid:2011egj}\\
$\mathcal{B}(B^0_s\to f_{2}(1270)\tau^+\tau^-)$                      &     $\cdots$       &     $2.11\pm 1.02$     &     $1.02\pm 0.49$     &     $...$ \\
$\mathcal{B}(B^0_s\to f^{\prime}_{2}(1525)\tau^+\tau^-)$             &     $\cdots$       &     $6.65\pm 2.46$     &     $3.21\pm 1.06$     &     $5.8^{+3.7}_{-2.1}$ \cite{Li:2010ra}, $1.36^{+0.41}_{-0.30}$ \cite{Zuo:2021kui}\\\hline
{\color{blue}$\overline{b}\to \overline{d}\ell^+\ell^-:$}\\
$\mathcal{B}(B^+_u\to a_{2}(1320)^+e^+e^-)$                          &     $0.12\pm 0.02$       &     $0.17\pm 0.07$     &     $0.15\pm 0.04$     &      $0.235^{+0.056}_{-0.047}$ \cite{Zuo:2021kui} \\
$\mathcal{B}(B^0_d\to a_{2}(1320)^{0}e^+e^-)$                        &     $0.056\pm 0.010$     &     $0.081\pm 0.033$     &     $0.069\pm 0.020$   &      $0.109^{+0.026}_{-0.022}$ \cite{Zuo:2021kui} \\
$\mathcal{B}(B^0_d\to f_{2}(1270)e^+e^-)$                            &     $0.061\pm 0.011$     &     $0.091\pm 0.035$   &     $0.079\pm 0.020$     &      $0.123^{+0.028}_{-0.024}$ \cite{Zuo:2021kui}\\
$\mathcal{B}(B^0_d\to f^{\prime}_{2}(1525)e^+e^-)$                   &     $(8.48\pm 3.11)\times 10^{-4}$    &     $(1.15\pm 0.67)\times 10^{-3}$     &  $(9.56\pm 4.06)\times 10^{-4}$     &      $...$ \\
$\mathcal{B}(B^0_s\to \overline{K}^{*}_{2}(1430)^0e^+e^-)$           &     $0.093\pm 0.017$     &     $0.14\pm 0.06$     &     $0.11\pm 0.03$     &       $0.182^{+0.055}_{-0.039}$ \cite{Zuo:2021kui} \\
$\mathcal{B}(B^+_u\to a_{2}(1320)^+\mu^+\mu^-)$                      &     $0.12\pm 0.02$       &     $0.13\pm 0.03$     &     $0.13\pm 0.02$     &     $0.128^{+0.031}_{-0.026}$ \cite{Zuo:2021kui} \\              		
$\mathcal{B}(B^0_d\to a_{2}(1320)^{0}\mu^+\mu^-)$                    &     $0.056\pm 0.010$     &     $0.058\pm 0.011$     &     $0.061\pm 0.012$     &     $0.059^{+0.015}_{-0.011}$ \cite{Zuo:2021kui} \\
$\mathcal{B}(B^0_d\to f_{2}(1270)\mu^+\mu^-)$                        &     $0.060\pm0.011$      &     $0.066\pm 0.014$    &    $0.068\pm0.013$     &     $0.065^{+0.016}_{-0.013}$ \cite{Zuo:2021kui} \\     	
$\mathcal{B}(B^0_d\to f^{\prime}_{2}(1525)\mu^+\mu^-)$               &     $(8.42\pm 3.10)\times 10^{-4}$       &     $(7.79\pm 2.71)\times 10^{-4}$     &     $(7.86\pm 2.93)\times 10^{-4}$     &     $...$ \\
$\mathcal{B}(B^0_s\to \overline{K}^{*}_{2}(1430)^0\mu^+\mu^-)$       &     $0.092\pm 0.017$     &     $0.10\pm 0.02$     &     $0.098\pm 0.018$     &     $0.108^{+0.032}_{-0.024}$ \cite{Zuo:2021kui}\\
$\mathcal{B}(B^+_u\to a_{2}(1320)^+\tau^+\tau^-)$                    &     $\cdots$       &     $1.27\pm 0.49$     &     $0.58\pm 0.19$     &     $0.219^{+0.053}_{-0.044}$ \cite{Zuo:2021kui} \\
$\mathcal{B}(B^0_d\to a_{2}(1320)^{0}\tau^+\tau^-)$                  &     $\cdots$       &     $0.59\pm 0.22$     &     $0.27\pm 0.09$     &     $0.102^{+0.023}_{-0.021}$ \cite{Zuo:2021kui} \\
$\mathcal{B}(B^0_d\to f_{2}(1270)\tau^+\tau^-)$                      &     $\cdots$       &     $0.88\pm 0.35$     &     $0.40\pm 0.14$     &     $0.161^{+0.037}_{-0.033}$ \cite{Zuo:2021kui}\\
$\mathcal{B}(B^0_d\to f^{\prime}_{2}(1525)\tau^+\tau^-)$             &     $\cdots$       &     $(1.07\pm 0.52)\times 10^{-3}$     &     $(5.27\pm 2.47)\times 10^{-4}$     &     $...$ \\
$\mathcal{B}(B^0_s\to \overline{K}^{*}_{2}(1430)^0\tau^+\tau^-)$     &     $\cdots$       &     $0.85\pm 0.33$     &     $0.39\pm 0.13$     &     $0.148^{+0.046}_{-0.033}$ \cite{Zuo:2021kui} \\\hline        	
\end{tabular}}
\end{center}\label{Tab:data1BrB2TllPred}
\end{table}

\subsubsection{Results in $S_{PQCD}$ scheme}
In order to obtain more precise observables, one also needs to consider the $q^2$ dependence of the form factors for the $B\to T\ell^+\ell^-$  decays.
In the $S_{\rm PQCD}$ scheme, the $q^2$ dependence of the $B\to T$ form factors is included by using the form-factor ratios obtained in the perturbative QCD approach  \cite{Li:2010ra}. The SU(3) relations in Tab. \ref{Tab:ARB2T} are applied to the normalization of $V(0)$, while the remaining form factors are written as $F_i(0)=r_i V(0)$,
and the values of the ratios $r_i\equiv\frac{F_i(0)}{V(0)}$ are taken from the PQCD results in Ref. \cite{Li:2010ra}. We obtain $V(0)=0.22\pm0.11$, which is similar to  the PQCD result $V(0)=0.20\pm0.04$ with larger error. The branching ratios of the $B \to T\ell^+\ell^-$ decays are predicted by the obtained $V(0)=0.22\pm0.11$, and the estimates are given in the third column of Tab. \ref{Tab:data1BrB2TllPred}. The larger uncertainties in this scheme mainly arise from the form factor ratios $r_i$ and their model dependence.

 Compared with $S_1$, this scheme allows us to calculate not only the integrated branching ratios but also the $q^2$-binned observables, such as $f_L$, $\overline{A}_{FB}$, $P_1$, $P_2$, $P_4'$, and $P_5'$,
and they are listed in Tab.  \ref{Tab:ObPQCDbin}. The observables of the $B\to T\tau^+\tau^-$ decays  are listed in Tab. \ref{Tab:ObPQCDtau}.

\begin{sidewaystable}
\centering
\renewcommand\arraystretch{1}\tabcolsep 0.15in
\caption{Observables  for the $B \to T\ell^+\ell^-$ decays in different $q^2$ bins  within  1$\sigma$ error in $S_{PQCD}$ scheme.  }
\resizebox{1.0\textwidth}{!}{
\begin{tabular}{lccccccc} \hline\hline
~~~~Observables                                                                   &         $[q^2_{min},2]$        &         $[2,4.3]$               &       $[4.3,8.68]$             &         $[10.09,12.86]$        & $ [14.18,q^2_{max}] $          &             $[1,6]$        &         $[q^2_{min},q^2_{max}]$    \\ \hline
$\mathcal{B}(B^+_u\to K^{*}_{2}(1430)^+e^+e^-)(\times 10^{-7})$               &         $1.65\pm 1.02$         &         $0.34\pm 0.11$          &       $0.68\pm 0.18$           &         $0.25\pm 0.06$         & $(3.04\pm 1.06)\times 10^{-3}$ &       $0.76\pm 0.23$        &         $2.90\pm 1.27$        \\
$\mathcal{B}(B^0_d\to \bar{K}^{*}_{2}(1430)^0e^+e^-)(\times 10^{-7})$         &         $1.54\pm 0.91$         &         $0.31\pm 0.10$          &       $0.63\pm 0.16$           &         $0.22\pm 0.06$         & $(2.47\pm 0.79)\times 10^{-3}$ &       $0.70\pm 0.21$        &         $2.68\pm 1.16$        \\
$\mathcal{B}(B^0_s\to f_{2}(1270)e^+e^-)(\times 10^{-7})$                     &         $0.058\pm 0.036$       &         $0.015\pm 0.007$        &       $0.032\pm 0.012$         &         $0.016\pm 0.006$       & $(3.32\pm 1.44)\times 10^{-3}$ &       $0.033\pm 0.015$      &         $0.13\pm 0.06$        \\
$\mathcal{B}(B^0_s\to f^{\prime}_{2}(1525)e^+e^-)(\times 10^{-7})$            &         $1.40\pm 0.84$         &         $0.28\pm 0.09$          &       $0.56\pm 0.14$           &         $0.21\pm 0.05$         & $(2.42\pm 0.98)\times 10^{-3}$ &       $0.63\pm 0.19$        &         $2.62\pm 0.83$        \\
$\mathcal{B}(B^+_u\to a_{2}(1320)^+e^+e^-)(\times 10^{-7})$                   &         $0.093\pm 0.055$       &         $0.020\pm 0.008$        &       $0.042\pm 0.012$         &         $0.019\pm 0.005$       & $(1.35\pm 0.39)\times 10^{-3}$ &       $0.046\pm 0.016$      &         $0.17\pm 0.07$        \\
$\mathcal{B}(B^0_d\to a_{2}(1320)^{0}e^+e^-)(\times 10^{-7})$                 &         $0.043\pm 0.026$       & $(9.43\pm 3.60)\times 10^{-3}$  &       $0.020\pm 0.005$         &        $0.0088\pm 0.0024$      & $(6.31\pm 1.81)\times 10^{-4}$ &       $0.021\pm 0.007$      &         $0.081\pm 0.033$      \\
$\mathcal{B}(B^0_d\to f_{2}(1270)e^+e^-)(\times 10^{-7})$                     &         $0.046\pm 0.027$       &         $0.010\pm 0.004$        &       $0.022\pm 0.006$         &         $0.010\pm 0.003$       & $(1.12\pm 0.33)\times 10^{-3}$ &       $0.023\pm 0.008$      &         $0.091\pm 0.035$      \\
$\mathcal{B}(B^0_d\to f^{\prime}_{2}(1525)e^+e^-)(\times 10^{-11})$           &         $7.15\pm 4.57$         &         $1.53\pm 0.67$          &       $2.92\pm 1.07$           &         $0.78\pm 0.32$         & $(2.07\pm 2.07)\times 10^{-10}$&       $3.39\pm 1.42$        &         $11.5\pm 6.66$        \\
$\mathcal{B}(B^0_s\to K^{*}_{2}(1430)^0e^+e^-)(\times 10^{-7})$               &         $0.076\pm 0.046$       &         $0.016\pm 0.006$        &       $0.033\pm 0.009$         &         $0.014\pm 0.004$       & $(7.88\pm 2.40)\times 10^{-4}$ &       $0.036\pm 0.012$      &         $0.14\pm 0.06$        \\ \hline
$\mathcal{B}(B^+_u\to K^{*}_{2}(1430)^+\mu^+\mu^-)(\times 10^{-7})$           &         $0.49\pm 0.21$         &         $0.34\pm 0.11$          &       $0.68\pm 0.18$           &         $0.25\pm 0.06$         & $(3.04\pm 1.07)\times 10^{-3}$ &       $0.76\pm 0.22$        &         $1.97\pm 0.34$        \\
$\mathcal{B}(B^0_d\to \bar{K}^{*}_{2}(1430)^0\mu^+\mu^-)(\times 10^{-7})$     &         $0.45\pm 0.19$         &         $0.31\pm 0.10$          &       $0.63\pm 0.16$           &         $0.22\pm 0.06$         & $(2.43\pm 0.78)\times 10^{-3}$ &       $0.69\pm 0.21$        &         $1.80\pm 0.30$        \\
$\mathcal{B}(B^0_s\to f_{2}(1270)\mu^+\mu^-)(\times 10^{-7})$                  &         $0.019\pm 0.009$       &         $0.015\pm 0.007$        &       $0.032\pm 0.012$         &         $0.016\pm 0.006$       & $(3.30\pm 1.42)\times 10^{-3}$ &       $0.033\pm 0.015$      &         $0.096\pm 0.033$      \\
$\mathcal{B}(B^0_s\to f^{\prime}_{2}(1525)\mu^+\mu^-)(\times 10^{-7})$        &         $0.41\pm 0.18$         &         $0.28\pm 0.09$          &       $0.56\pm 0.14$           &         $0.21\pm 0.05$         & $(2.40\pm 0.97)\times 10^{-3}$ &       $0.62\pm 0.18$        &         $1.62\pm 0.22$        \\
$\mathcal{B}(B^+_u\to a_{2}(1320)^+\mu^+\mu^-)(\times 10^{-7})$               &         $0.027\pm 0.012$       &         $0.020\pm 0.008$        &       $0.042\pm 0.012$         &         $0.019\pm 0.005$       & $(1.35\pm 0.39)\times 10^{-3}$ &       $0.046\pm 0.016$      &         $0.13\pm 0.03$        \\
$\mathcal{B}(B^0_d\to a_{2}(1320)^{0}\mu^+\mu^-)(\times 10^{-7})$             &         $0.013\pm 0.005$       &  $(9.48\pm 3.50)\times 10^{-3}$ &       $0.019\pm 0.005$         &        $0.0087\pm 0.0024$      & $(6.31\pm 1.83)\times 10^{-4}$ &       $0.021\pm 0.007$      &         $0.058\pm 0.011$      \\
$\mathcal{B}(B^0_d\to f_{2}(1270)\mu^+\mu^-)(\times 10^{-7})$                 &         $0.013\pm 0.006$       &         $0.010\pm 0.004$        &       $0.022\pm 0.006$         &         $0.010\pm 0.003$       & $(1.13\pm 0.34)\times 10^{-3}$ &       $0.023\pm 0.008$      &         $0.066\pm 0.014$      \\
$\mathcal{B}(B^0_d\to f^{\prime}_{2}(1525)\mu^+\mu^-)(\times 10^{-11})$       &         $2.19\pm 1.07$         &         $1.53\pm 0.66$          &       $2.90\pm 1.06$           &         $0.78\pm 0.32$         & $(2.05\pm 2.05)\times 10^{-10}$&       $3.38\pm 1.41$        &         $7.79\pm 2.71$        \\
$\mathcal{B}(B^0_s\to K^{*}_{2}(1430)^0\mu^+\mu^-) $                          &         $0.022\pm 0.010$       &         $0.016\pm 0.006$        &       $0.032\pm 0.009$         &         $0.014\pm 0.004$       & $(7.79\pm 2.43)\times 10^{-4}$ &       $0.036\pm 0.011$      &         $0.098\pm 0.018$      \\ \hline
$f_L(B^+_u\to K^{*}_{2}(1430)^+e^+e^-) $               &         $0.29\pm 0.24$         &         $0.77\pm 0.15$          &       $0.63\pm 0.15$           &         $0.47\pm 0.10$         &    $0.40\pm 0.04$              &       $0.74\pm 0.17$        &         $0.49\pm 0.28$        \\
$A_{FB}(B^+_u\to K^{*}_{2}(1430)^+e^+e^-) $               &         $0.040\pm 0.012$       &         $0.016\pm 0.124$        &       $-0.20\pm 0.11$          &         $-0.32\pm 0.13$        &      $-0.23\pm 0.11$           &       $-0.016\pm 0.118$      &         $-0.11\pm 0.09$            \\
$P_1(B^+_u\to K^{*}_{2}(1430)^+e^+e^-) $               &         $-0.0099\pm 0.4410$     &         $-0.051\pm 0.463$        &       $-0.11\pm 0.54$          &         $-0.31\pm 0.45$        &      $-0.80\pm 0.18$           &       $-0.071\pm 0.440$      &         $-0.083\pm 0.399$            \\
$P_2(B^+_u\to K^{*}_{2}(1430)^+e^+e^-) $               &         $-0.035\pm 0.008$      &         $-0.027\pm 0.314$       &       $0.37\pm 0.10$           &         $0.39\pm 0.10$         &    $0.25\pm 0.12$              &       $0.050\pm 0.257$      &         $0.13\pm 0.11$           \\
$P'_4(B^+_u\to K^{*}_{2}(1430)^+e^+e^-) $               &         $-0.18\pm 0.10$        &         $0.74\pm 0.37$          &       $0.98\pm 0.29$           &         $1.13\pm 0.22$         &    $1.36\pm 0.10$              &       $0.70\pm 0.35$        &         $0.49\pm 0.25$           \\
$P'_5(B^+_u\to K^{*}_{2}(1430)^+e^+e^-) $               &         $0.23\pm 0.12$         &         $-0.59\pm 0.44$         &       $-0.83\pm 0.33$          &         $-0.76\pm 0.30$        &      $-0.39\pm 0.19$           &       $-0.56\pm 0.42$        &         $-0.36\pm 0.28$            \\\hline
$f_L(B^+_u\to K^{*}_{2}(1430)^+\mu^+\mu^-) $           &         $0.52\pm 0.32$         &         $0.77\pm 0.14$          &       $0.63\pm 0.15$           &         $0.47\pm 0.10$         &    $0.40\pm 0.03$              &       $0.75\pm 0.16$        &         $0.61\pm 0.18$        \\
$A_{FB}(B^+_u\to K^{*}_{2}(1430)^+\mu^+\mu^-) $           &         $0.098\pm 0.056$       &         $0.015\pm 0.122$        &       $-0.20\pm 0.11$          &         $-0.32\pm 0.13$        &      $-0.23\pm 0.11$           &       $-0.018\pm 0.115$      &         $-0.14\pm 0.10$            \\
$P_1(B^+_u\to K^{*}_{2}(1430)^+\mu^+\mu^-) $           &         $-0.012\pm 0.464$       &         $-0.048\pm 0.461$        &       $-0.11\pm 0.53$          &         $-0.31\pm 0.45$        &      $-0.81\pm 0.17$           &       $-0.071\pm 0.440$      &         $-0.18\pm 0.42$        \\
$P_2(B^+_u\to K^{*}_{2}(1430)^+\mu^+\mu^-) $           &         $-0.20\pm 0.05$        &         $-0.026\pm 0.316$       &       $0.37\pm 0.11$           &         $0.39\pm 0.09$         &    $0.26\pm 0.12$              &       $0.053\pm 0.259$       &         $0.22\pm 0.15$           \\
$P'_4(B^+_u\to K^{*}_{2}(1430)^+\mu^+\mu^-) $           &         $-0.20\pm 0.21$        &         $0.74\pm 0.38$          &       $0.98\pm 0.28$           &         $1.13\pm 0.22$         &    $1.34\pm 0.09$              &       $0.70\pm 0.35$       &         $0.74\pm 0.28$           \\
$P'_5(B^+_u\to K^{*}_{2}(1430)^+\mu^+\mu^-) $           &         $0.36\pm 0.27$         &         $-0.59\pm 0.45$         &       $-0.83\pm 0.32$          &         $-0.76\pm 0.30$        &      $-0.39\pm 0.19$           &       $-0.57\pm 0.43$        &         $-0.55\pm 0.36$         \\\hline
\end{tabular}}	\label{Tab:ObPQCDbin}
\end{sidewaystable}

\begin{sidewaystable}
\centering
\renewcommand\arraystretch{1}\tabcolsep 0.15in
\caption{Observables  for the $B \to T\ell^+\ell^-$ decays in different $q^2$ bins  within  1$\sigma$ error in $S_{LCSR}$ scheme.  }
\resizebox{1.0\textwidth}{!}{
\begin{tabular}{lccccccc} \hline\hline
Observables                                                                   &         $[q^2_{min},2]$        &         $[2,4.3]$               &       $[4.3,8.68]$             &         $[10.09,12.86]$        & $ [14.18,q^2_{max}] $          &             $[1,6]$        &         $[q^2_{min},q^2_{max}]$    \\ \hline
$\mathcal{B}(B^+_u\to K^{*}_{2}(1430)^+e^+e^-)(\times 10^{-7})$               &         $1.01\pm 0.42$         &         $0.52\pm 0.10$          &       $0.68\pm 0.14$           &         $0.12\pm 0.03$         & $(9.41\pm 2.40)\times 10^{-4}$ &          $1.10\pm 0.22$    &         $2.38\pm 0.64$        \\
$\mathcal{B}(B^0_d\to \bar{K}^{*}_{2}(1430)^0e^+e^-)(\times 10^{-7})$         &         $0.95\pm 0.36$         &         $0.48\pm 0.09$          &       $0.62\pm 0.12$           &         $0.11\pm 0.02$         & $(7.66\pm 1.91)\times 10^{-4}$ &          $0.99\pm 0.19$    &         $2.18\pm 0.60$        \\
$\mathcal{B}(B^0_s\to f_{2}(1270)e^+e^-)(\times 10^{-7})$                     &         $0.041\pm 0.020$       &         $0.023\pm 0.009$        &       $0.033\pm 0.012$         &         $0.0078\pm 0.0029$     & $(9.36\pm 3.54)\times 10^{-4}$ &          $0.049\pm 0.018$  &         $0.11\pm 0.04$        \\
$\mathcal{B}(B^0_s\to f^{\prime}_{2}(1525)e^+e^-)(\times 10^{-7})$            &         $0.86\pm 0.32$         &         $0.41\pm 0.07$          &       $0.55\pm 0.11$           &         $0.098\pm 0.021$       & $(7.87\pm 2.84)\times 10^{-4}$ &          $0.87\pm 0.15$    &         $2.03\pm 0.41$        \\
$\mathcal{B}(B^+_u\to a_{2}(1320)^+e^+e^-)(\times 10^{-7})$                   &         $0.062\pm 0.025$       &         $0.033\pm 0.007$        &       $0.044\pm 0.009$         &         $0.0089\pm 0.0020$     & $(4.02\pm 0.91)\times 10^{-4}$ &          $0.069\pm 0.015$  &         $0.15\pm 0.04$        \\
$\mathcal{B}(B^0_d\to a_{2}(1320)^{0}e^+e^-)(\times 10^{-7})$                 &         $0.029\pm 0.011$       &         $0.015\pm 0.003$        &       $0.021\pm 0.004$         &         $0.0041\pm 0.0009$     & $(1.86\pm 0.41)\times 10^{-4}$ &          $0.032\pm 0.007$  &         $0.069\pm 0.020$        \\
$\mathcal{B}(B^0_d\to f_{2}(1270)e^+e^-)(\times 10^{-7})$                     &         $0.030\pm 0.012$       &         $0.017\pm 0.004$        &       $0.023\pm 0.005$         &         $0.0049\pm 0.0011$     & $(3.23\pm 0.72)\times 10^{-4}$ &          $0.036\pm 0.008$  &         $0.079\pm 0.020$        \\
$\mathcal{B}(B^0_d\to f^{\prime}_{2}(1525)e^+e^-)(\times 10^{-11})$           &         $4.45\pm 2.23$         &         $2.12\pm 0.81$          &       $2.66\pm 0.97$           &         $0.38\pm 0.15$         & $(9.32\pm 9.32)\times 10^{-7}$ &          $4.39\pm 1.68$    &         $9.56\pm 4.06$        \\
$\mathcal{B}(B^0_s\to K^{*}_{2}(1430)^0e^+e^-)(\times 10^{-7})$               &         $0.047\pm 0.020$       &         $0.025\pm 0.005$        &       $0.033\pm 0.007$         &         $0.0067\pm 0.0015$     & $(2.34\pm 0.56)\times 10^{-4}$ &          $0.052\pm 0.011$  &         $0.11\pm 0.03$        \\ \hline
$\mathcal{B}(B^+_u\to K^{*}_{2}(1430)^+\mu^+\mu^-)(\times 10^{-7})$           &         $0.59\pm 0.16$         &         $0.52\pm 0.10$          &       $0.68\pm 0.14$           &         $0.12\pm 0.03$         & $(9.48\pm 2.44)\times 10^{-4}$ &          $1.08\pm 0.20$    &         $2.01\pm 0.35$        \\
$\mathcal{B}(B^0_d\to \bar{K}^{*}_{2}(1430)^0\mu^+\mu^-)(\times 10^{-7})$     &         $0.54\pm 0.14$         &         $0.47\pm 0.09$          &       $0.62\pm 0.13$           &         $0.10\pm 0.02$         & $(7.58\pm 1.86)\times 10^{-4}$ &          $1.00\pm 0.18$    &         $1.83\pm 0.31$        \\
$\mathcal{B}(B^0_s\to f_{2}(1270)\mu^+\mu^-)(\times 10^{-7})$                  &         $0.024\pm 0.014$       &         $0.023\pm 0.008$        &       $0.033\pm 0.012$         &         $0.0079\pm 0.0029$     & $(9.43\pm 3.60)\times 10^{-4}$ &          $0.048\pm 0.018$  &         $0.096\pm 0.033$        \\
$\mathcal{B}(B^0_s\to f^{\prime}_{2}(1525)\mu^+\mu^-)(\times 10^{-7})$        &         $0.47\pm 0.12$         &         $0.41\pm 0.07$          &       $0.55\pm 0.11$           &         $0.098\pm 0.021$       & $(7.90\pm 2.80)\times 10^{-4}$ &          $0.86\pm 0.15$    &         $1.62\pm 0.22$        \\
$\mathcal{B}(B^+_u\to a_{2}(1320)^+\mu^+\mu^-)(\times 10^{-7})$               &         $0.036\pm 0.010$       &         $0.033\pm 0.007$        &       $0.044\pm 0.009$         &         $0.0089\pm 0.0019$     & $(3.99\pm 0.91)\times 10^{-4}$ &          $0.069\pm 0.015$  &         $0.13\pm 0.02$        \\
$\mathcal{B}(B^0_d\to a_{2}(1320)^{0}\mu^+\mu^-)(\times 10^{-7})$             &         $0.017\pm 0.004$       &         $0.015\pm 0.003$        &       $0.021\pm 0.004$         &         $0.0041\pm 0.0009$     & $(1.85\pm 0.41)\times 10^{-4}$ &          $0.032\pm 0.007$  &         $0.061\pm 0.012$        \\
$\mathcal{B}(B^0_d\to f_{2}(1270)\mu^+\mu^-)(\times 10^{-7})$                 &         $0.018\pm 0.005$       &         $0.017\pm 0.004$        &       $0.023\pm 0.005$         &         $0.0049\pm 0.0011$     & $(3.25\pm 0.77)\times 10^{-4}$ &          $0.036\pm 0.008$  &         $0.068\pm 0.013$        \\
$\mathcal{B}(B^0_d\to f^{\prime}_{2}(1525)\mu^+\mu^-)(\times 10^{-11})$       &         $2.40\pm 0.95$         &         $2.10\pm 0.78$          &       $2.64\pm 0.98$           &         $0.38\pm 0.14$         & $(9.34\pm 9.34)\times 10^{-7}$ &          $4.39\pm 1.66$    &         $7.86\pm 2.93$        \\
$\mathcal{B}(B^0_s\to K^{*}_{2}(1430)^0\mu^+\mu^-)(\times 10^{-7})$           &         $0.027\pm 0.007$       &         $0.025\pm 0.005$        &       $0.034\pm 0.007$         &         $0.0067\pm 0.0015$     & $(2.32\pm 0.55)\times 10^{-4}$ &          $0.052\pm 0.011$  &         $0.098\pm 0.019$        \\ \hline
$f_L(B^+_u\to K^{*}_{2}(1430)^+e^+e^-)$                                       &         $0.62\pm 0.30$         &         $0.94\pm 0.03$          &         $0.83\pm 0.04$         &         $0.59\pm 0.03$         &          $0.43\pm 0.02$         &          $0.92\pm 0.04$        &         $0.73\pm 0.16$        \\
$A_{FB}(B^+_u\to K^{*}_{2}(1430)^+e^+e^-)$                                    &         $0.028\pm 0.009$      &        $-0.0050\pm 0.0189$      &       $-0.094\pm 0.032$         &         $-0.19\pm 0.06$       &          $-0.11\pm 0.03$       &          $-0.011\pm 0.019$      &         $-0.038\pm 0.022$    \\
$P_1(B^+_u\to K^{*}_{2}(1430)^+e^+e^-)$                                       &         $0.0020\pm 0.2940$     &         $-0.43\pm 0.24$         &       $-0.54\pm 0.23$          &         $-0.73\pm 0.14$        &          $-0.97\pm 0.06$        &          $-0.42\pm 0.24$       &         $-0.19\pm 0.24$       \\
$P_2(B^+_u\to K^{*}_{2}(1430)^+e^+e^-)$                                       &         $-0.038\pm 0.005$      &         $0.056\pm 0.175$        &         $0.33\pm 0.08$         &         $0.31\pm 0.08$         &          $0.13\pm 0.04$         &          $0.093\pm 0.147$      &         $0.089\pm 0.059$    \\
$P'_4(B^+_u\to K^{*}_{2}(1430)^+e^+e^-)$                                      &         $-0.13\pm 0.08$        &         $1.02\pm 0.19$          &         $1.19\pm 0.12$         &         $1.31\pm 0.08$         &          $1.41\pm 0.05$         &          $0.94\pm 0.20$        &         $0.51\pm 0.19$          \\
$P'_5(B^+_u\to K^{*}_{2}(1430)^+e^+e^-)$                                      &         $0.27\pm 0.08$         &         $-0.41\pm 0.23$         &       $-0.58\pm 0.18$          &         $-0.48\pm 0.14$        &          $-0.19\pm 0.06$        &          $-0.35\pm 0.22$       &         $-0.13\pm 0.12$        \\ \hline
$f_L(B^+_u\to K^{*}_{2}(1430)^+\mu^+\mu^-)$                                   &         $0.81\pm 0.12$         &         $0.94\pm 0.03$          &         $0.82\pm 0.04$         &         $0.59\pm 0.03$         &          $0.42\pm 0.02$         &          $0.92\pm 0.04$        &         $0.83\pm 0.06$        \\
$A_{FB}(B^+_u\to K^{*}_{2}(1430)^+\mu^+\mu^-)$                                &         $0.043\pm 0.020$      &        $-0.0050\pm 0.0189$      &       $-0.094\pm 0.032$         &         $-0.19\pm 0.05$       &          $-0.11\pm 0.04$       &          $-0.011\pm 0.019$      &         $-0.046\pm 0.023$    \\
$P_1(B^+_u\to K^{*}_{2}(1430)^+\mu^+\mu^-)$                                   &         $0.013\pm 0.311$       &         $-0.43\pm 0.24$         &       $-0.54\pm 0.23$          &         $-0.73\pm 0.15$        &          $-0.96\pm 0.06$        &          $-0.42\pm 0.23$       &         $-0.46\pm 0.21$       \\
$P_2(B^+_u\to K^{*}_{2}(1430)^+\mu^+\mu^-)$                                   &         $-0.22\pm 0.03$        &         $0.057\pm 0.176$        &         $0.33\pm 0.08$         &         $0.31\pm 0.08$         &          $0.13\pm 0.04$         &          $0.097\pm 0.147$      &         $0.18\pm 0.08$    \\
$P'_4(B^+_u\to K^{*}_{2}(1430)^+\mu^+\mu^-)$                                  &         $-0.080\pm 0.187$      &         $1.03\pm 0.18$          &         $1.20\pm 0.11$         &         $1.31\pm 0.07$         &          $1.40\pm 0.06$         &          $0.95\pm 0.19$        &         $0.82\pm 0.18$          \\
$P'_5(B^+_u\to K^{*}_{2}(1430)^+\mu^+\mu^-)$                                  &         $0.46\pm 0.16$         &         $-0.41\pm 0.23$         &       $-0.59\pm 0.19$          &         $-0.48\pm 0.14$        &          $-0.19\pm 0.06$        &          $-0.36\pm 0.22$       &         $-0.27\pm 0.18$ \\\hline
\end{tabular}}\label{Tab:ObLCSRbin}
\end{sidewaystable}

\begin{table}[t]
\centering\renewcommand\arraystretch{1.15}\tabcolsep 0.08in
\caption{Observables for the $B \to T\tau^+\tau^-$ decays  within  1$\sigma$ error in $S_{PQCD}$ scheme.  }
\begin{tabular}{lcccccccc}\hline\hline
~~~~~~Decays modes                                         &       $f_L$            &   $A_{FB}$            &       $P_1$            &     $P_2$          &  $P'_4$            &  $P'_5$              \\ \hline
$B^+_u\to K^{*}_{2}(1430)^+\tau^+\tau^-$             &  $0.58\pm 0.12$        &   $-0.097\pm0.041$    &   $-0.63\pm0.29$       & $1.29\pm0.51$      & $1.27\pm0.14$      & $-2.13\pm0.98$       \\			
$B^0_d\to \bar{K}^{*}_{2}(1430)^0\tau^+\tau^-$       &  $0.57\pm 0.11$        &   $-0.096\pm0.041$    &   $-0.63\pm0.29$       & $1.28\pm0.50$      & $1.28\pm0.15$      & $-2.16\pm1.00$       \\
$B^0_s\to f_{2}(1270)\tau^+\tau^-$                    &  $0.60\pm 0.13$        &   $-0.139\pm0.054$    &   $-0.42\pm0.41$       & $1.09\pm0.31$      & $1.18\pm0.20$      & $-2.06\pm0.85$       \\
$B^0_s\to f^{\prime}_{2}(1525)\tau^+\tau^-$          &  $0.58\pm 0.12$        &   $-0.095\pm0.041$    &   $-0.64\pm0.28$       & $1.27\pm0.50$      & $1.28\pm0.14$      & $-2.10\pm0.97$       \\
$B^+_u\to a_{2}(1320)^+\tau^+\tau^-$                 &  $0.58\pm 0.13$        &   $-0.12\pm0.05$      &   $-0.52\pm0.36$       & $1.20\pm0.41$      & $1.23\pm0.17$      & $-2.12\pm0.93$      \\
$B^0_d\to a_{2}(1320)^{0}\tau^+\tau^-$               &  $0.59\pm 0.12$        &   $-0.12\pm0.05$      &   $-0.53\pm0.36$       & $1.18\pm0.40$      & $1.23\pm0.16$      & $-2.12\pm0.93$      \\
$B^0_d\to f_{2}(1270)\tau^+\tau^-$                   &  $0.59\pm 0.13$        &   $-0.12\pm0.05$      &   $-0.48\pm0.38$       & $1.16\pm0.37$      & $1.21\pm0.17$      & $-2.10\pm0.91$      \\
$B^0_d\to f^{\prime}_{2}(1525)\tau^+\tau^-$          &  $0.55\pm 0.10$        &   $-0.077\pm0.036$    &   $-0.73\pm0.23$       & $1.36\pm0.59$      & $1.32\pm0.10$      & $-2.18\pm1.04$      \\
$B^0_s\to K^{*}_{2}(1430)^0\tau^+\tau^-$             &  $0.59\pm 0.13$        &   $-0.11\pm0.05$      &   $-0.56\pm0.34$       & $1.20\pm0.43$      & $1.25\pm0.16$      & $-2.11\pm0.95$      \\ \hline
\end{tabular}\label{Tab:ObPQCDtau} \vspace{1cm}
%\end{table}
%\begin{table}[t]
\centering\renewcommand\arraystretch{1.15}\tabcolsep 0.08in
\caption{Observables for the $B \to T\tau^+\tau^-$ decays  within  1$\sigma$ error in $S_{LCSR}$ scheme.  }
\begin{tabular}{lcccccccc}\hline\hline
~~~~~~Decays modes                                     &       $f_L$       &   $A_{FB}$            &    $P_1$               &     $P_2$          &  $P'_4$            &  $P'_5$              \\ \hline
$B^+_u\to K^{*}_{2}(1430)^+\tau^+\tau^-$             &  $0.66\pm 0.08$   &   $-0.042\pm0.014$    &   $-0.92\pm0.07$       & $0.78\pm0.24$      & $1.40\pm0.07$      & $-1.16\pm0.35$       \\			
$B^0_d\to \bar{K}^{*}_{2}(1430)^0\tau^+\tau^-$       &  $0.65\pm 0.09$   &   $-0.042\pm0.014$    &   $-0.90\pm0.07$       & $0.79\pm0.23$      & $1.39\pm0.07$      & $-1.16\pm0.37$       \\
$B^0_s\to f_{2}(1270)\tau^+\tau^-$                    &  $0.74\pm 0.09$   &   $-0.052\pm0.016$    &   $-0.83\pm0.11$       & $0.76\pm0.21$      & $1.35\pm0.06$      & $-1.19\pm0.34$       \\
$B^0_s\to f^{\prime}_{2}(1525)\tau^+\tau^-$          &  $0.66\pm 0.09$   &   $-0.041\pm0.013$    &   $-0.91\pm0.06$       & $0.76\pm0.23$      & $1.39\pm0.06$      & $-1.14\pm0.35$       \\
$B^+_u\to a_{2}(1320)^+\tau^+\tau^-$                 &  $0.70\pm 0.08$   &   $-0.049\pm0.015$    &   $-0.87\pm0.09$       & $0.77\pm0.22$      & $1.37\pm0.06$      & $-1.18\pm0.34$      \\
$B^0_d\to a_{2}(1320)^{0}\tau^+\tau^-$               &  $0.70\pm 0.09$   &   $-0.049\pm0.015$    &   $-0.87\pm0.08$       & $0.77\pm0.22$      & $1.37\pm0.06$      & $-1.19\pm0.35$      \\
$B^0_d\to f_{2}(1270)\tau^+\tau^-$                   &  $0.72\pm 0.09$   &   $-0.051\pm0.016$    &   $-0.85\pm0.10$       & $0.78\pm0.21$      & $1.36\pm0.06$      & $-1.20\pm0.35$      \\
$B^0_d\to f^{\prime}_{2}(1525)\tau^+\tau^-$          &  $0.61\pm 0.08$   &   $-0.034\pm0.011$    &   $-0.94\pm0.05$       & $0.77\pm0.24$      & $1.39\pm0.06$      & $-1.14\pm0.35$      \\
$B^0_s\to K^{*}_{2}(1430)^0\tau^+\tau^-$             &  $0.69\pm 0.09$   &   $-0.047\pm0.015$    &   $-0.89\pm0.08$       & $0.76\pm0.22$      & $1.38\pm0.07$      & $-1.17\pm0.35$      \\ \hline
\end{tabular}\label{Tab:ObLCSRtau}
\end{table}

\subsubsection{Results in $S_{LCSR}$ scheme}

In the $S_{\rm LCSR}$ scheme, we follow the same normalization strategy as in the $S_{\rm PQCD}$ scheme, but use the form factors obtained from the LCSR \cite{Yang:2010qd}. This provides an independent treatment of the hadronic dynamics and allows us to estimate the theoretical uncertainty associated with different form factor inputs. The extracted value of $V(0)=0.13\pm0.05$  is consistent with the LCSR calculation result $V(0)=0.15\pm0.02$ within uncertainties, which supports the reliability of the normalization procedure. Differences between the $S_{\rm PQCD}$ and $S_{\rm LCSR}$ estimates therefore mainly originate from the different $q^2$ behavior of the corresponding form factors. The branching ratio estimates  of the $B \to T\ell^+\ell^-$ decays are listed  in the fourth column of Tab. \ref{Tab:data1BrB2TllPred}.
Other results are listed in Tab. \ref{Tab:ObLCSRbin} and Tab. \ref{Tab:ObLCSRtau}.

\subsubsection{Comparative analysis}
Comparing the three schemes, we find that most branching-ratio estimates for the $e^+e^-$ and $\mu^+\mu^-$ modes are consistent with each other within the $1\sigma$ uncertainties. This indicates that the SU(3) amplitude relations provide stable estimates once the overall normalization is fixed by the measured $B_s^0\to f_2'(1525)\mu^+\mu^-$ mode. For the $\tau^+\tau^-$ modes, the estimates in the $S_{\rm LCSR}$ scheme are generally smaller than those in the $S_{\rm PQCD}$ scheme.
This difference is likely related to the different $q^2$ dependence of the form factors adopted in the \(S_{\rm PQCD}\) and \(S_{\rm LCSR}\) schemes. Since the \(\tau^+\tau^-\) channels are kinematically restricted to a relatively high-\(q^2\) region, they are expected to be more sensitive to the form factor shapes than the \(e^+e^-\) and \(\mu^+\mu^-\) modes. Therefore, the comparison between $S_{\rm PQCD}$ and $S_{\rm LCSR}$ provides a useful estimate of the form factor induced theoretical uncertainty.

Previous results are also  listed in the last column of
Tab. \ref{Tab:data1BrB2TllPred}, and they have been  obtained by the
LCSR  \cite{Aliev:2019ojc}, the PQCD  \cite{Li:2010ra}, and the LCSR in the framework of heavy quark effective field theory (HQEFT) \cite{Zuo:2021kui}.
 Many of our estimates  are in good agreement with them. A few of our estimates, such as the branching ratios of the $B \to T \tau^+\tau^-$ decays with the $b\to d\tau^+\tau^-$ transition,  are slightly larger than those obtained in Ref. \cite{Zuo:2021kui}.  In addition, $B(B_s^0\to f_2(1270)\ell^+\ell^-)$ and $B(B_d^0\to f'_2(1525)\ell^+\ell^-)$ are given for the first time to our knowledge.

 For the uncertainties in the numerical results, they are from  the uncertainties of the experimental measurement of $\mathcal{B}(B^0_s\to f^{\prime}_{2}(1525)\mu^+\mu^-)$,  the CKM matrix elements, the $B$ meson lifetime, and the particle masses in three cases, and they also come from the form factor ratios $r_i$ in $S_{PQCD}$ and  $S_{LCSR}$ cases.   Comparing with  $H_{B\to T}=24.70\pm2.34$ in $S_1$ case, $V(0)=0.22\pm0.11$ in $S_{PQCD}$ case, and $V(0)=0.13\pm0.05$ in $S_{LCSR}$ case, their errors are $9.5\%$, $50\%$, and $38.5\%$. So the main uncertainty of the results is from the experimental measurement of $\mathcal{B}(B^0_s\to f^{\prime}_{2}(1525)\mu^+\mu^-)$ in $S_1$ case, and the main uncertainty of the results comes from  the uncertainties in the ratio $r_i$ in $S_{PQCD}$ and  $S_{LCSR}$ cases.

We also note that  $V(0)$ in the $S_{QCD}$ and $S_{LCSR}$ schemes is extracted from a single measured branching ratio. Since the branching ratios scale approximately as $|V(0)|^2$, the uncertainty of $V(0)$ is amplified in the branching ratio estimates. In addition, the possible SU(3) breaking effects have not been included in the quoted numerical errors because they cannot be reliably constrained with the currently available data. Thus, the uncertainties shown in the tables should be understood as those propagated from the input parameters and form factor ratios within each scheme, rather than complete theoretical uncertainties.

The three numerical schemes should therefore be regarded as phenomenological scenarios rather than three independent SU(3) predictions. The $S_1$ case is used as a symmetry based normalization benchmark, while the $S_{PQCD}$ and $S_{LCSR}$ schemes incorporate external information on the form factor ratios and their $q^2$ dependences. Therefore, the spread between the $S_{PQCD}$ and $S_{LCSR}$ results reflects the present model dependence of the hadronic inputs.

\subsubsection{Other observables}
In addition, the branching ratios, the longitudinal polarization fractions, the normalized forward-backward asymmetries,  and some other observables   with $\ell=e,\mu$ and the different $q^2$ bins are also obtained,  and they are listed in Tab. \ref{Tab:ObPQCDbin}  and Tab. \ref{Tab:ObLCSRbin} for  $S_{PQCD}$ case and  $S_{LCSR}$ case, respectively.
The polarization fractions and normalized observables are less sensitive to the overall normalization fixed from the measured branching fraction. They mainly probe the relative sizes and $q^2$ dependences of the form factors.
Thus, in the $S_{PQCD}$  and $S_{LCSR}$  cases, these observables should be understood as model dependent results constrained by the SU(3) flavor relations among different channels.
For $f_L$, $A_{FB}$, $P_{1,2}$ and $P'_{4,5}$, the results of different decay modes are similar to each other, so we only take $B^+_u\to K^*_2(1430)e^+e^-$ and  $B^+_u\to K^*_2(1430)\mu^+\mu^-$ as examples.  One can see that all observables  are  obviously different between $\ell=e$ and  $\ell=\mu$ in $q^2\in[q^2_{min},2]$, but they are almost the same in other $q^2$ bins.  Comparing the  observables in $S_{PQCD}$ and $S_{LCSR}$ cases,
one can see that the estimates are quite different
between $S_{PQCD}$ and $S_{LCSR}$ cases in all $q^2$ bins except $[4.3,8.68]$, which are mainly due to
the $q^2$ dependence of the hadronic amplitudes.

For the observables with $\ell=\tau$, since the $q^2$ ranges are too small for all decays,  we do not  divide the $q^2$ range into bins.
The $f_L$, $A_{FB}$, $P_{1,2}$ and $P'_{4,5}$ with $\ell=\tau$ over the whole $q^2$ range are given in Tab. \ref{Tab:ObPQCDtau} for the $S_{PQCD}$ case and in Tab. \ref{Tab:ObLCSRtau} for the $S_{LCSR}$ case.

%\clearpage
\section{Charmless  decays $B\to PP/PV\ell^+\ell^-$  with the tensor resonances} \label{sec:D4MM}

\subsection{ $\mathcal{B}(B\to PP/PV\ell^+\ell^-)$ with the narrow width approximation }
For the four body decays $B\to M_1M_2\ell^+\ell^-$, the following hadronic intermediate states  might contribute
$B\to R(\to M_1M_2)\ell^+\ell^-$, $B\to R(\to \ell^+\ell^-) M_1M_2$,  $B\to R(\to M_1\ell^+\ell^-) M_2$, and $B\to J/\psi(\to \ell^+\ell^-)R(\to M_1M_2) $,  here $R$ denotes a resonance.
In this work, we only consider the $B\to R(\to M_1M_2)\ell^+\ell^-$ with $R=T$.
For the $B\to MM\ell^+\ell^-$  decays with the resonances, if the
decay widths of the resonance states are very narrow, the resonance branching ratios can be obtained by the narrow width approximation
\begin{eqnarray}
\mathcal{B}(B\to T(\to M_1M_2)\ell^+\ell^-)=\mathcal{B}(B\to T\ell^+\ell^-)\mathcal{B}(T\to M_1M_2). \label{Eq:NWA3}
\end{eqnarray}
 The quantity in Eq. (\ref{Eq:NWA3}) should be interpreted as the contribution from a specific tensor resonance component, rather than the full four body decay ratio. In realistic $B\to M_1M_2\ell^+\ell^-$ decays, scalar, vector, axial vector, tensor, and nonresonant contributions may enter the same final state. Since tensor mesons carry $J^P=2^+$, their contributions can in principle be separated from other partial waves through an amplitude analysis based on the $M_1M_2$ invariant-mass distribution and the angular distributions of the final state particles. Therefore, the branching ratios obtained below provide estimates of the tensor resonance components that may be used as inputs for future partial wave analyses.
Firstly, we will only use the  narrow width approximation in Eq. (\ref{Eq:NWA3}) to obtain the branching ratios of the $B\to T(\to M_1M_2)\ell^+\ell^-$ decays in this work.

$\mathcal{B}(T \to P V )$ and $\mathcal{B}(T \to P P)$  have been studied in our previous works \cite{Qiao:2024nbq,RMWB2PPlv}. Using the expressions of $\mathcal{B}(B \to T \ell^+\ell^-)$ in $S_{PQCD}$ scheme,  $\mathcal{B}(T \to PP,PV )$ given in Refs. \cite{Qiao:2024nbq,RMWB2PPlv}, and relevant experimental data given in PDG \cite{PDG2025}, we can give the branching ratios of the $B\to T(\to PP)\ell^+\ell^-$ and $B\to T(\to PV)\ell^+\ell^-$ decays, and
they  are listed in  Tabs. \ref{Tab:data1PPllTbs}-\ref{Tab:data1PPllTbd}  and Tabs. \ref{Tab:data1PVllTbs}-\ref{Tab:data1PVllTbd}, respectively.

\begin{table}[htb]
\caption{Branching ratios for the $B\to PP\ell^+\ell^-$ decays with the tensor  resonances  under the $b\to s \ell^+\ell^-$ transitions within  1$\sigma$ error via the narrow width approximation. Ones for $\ell=e,\mu$ are  in unit of $10^{-7}$, and ones for $\ell=\tau$ are in unit of $10^{-9}$. }
\centering{\footnotesize
\begin{center}  \renewcommand\arraystretch{1.05}\tabcolsep0.3in
\begin{tabular}{lccc} \hline\hline
                                 ~~~~~~~~~~  Decay modes   &       $\mathcal{B}$ with  $\ell=e$        &      $\mathcal{B}$ with   $\ell=\mu$     &    $\mathcal{B}$  with $\ell=\tau$     \\ \hline
$B^+_u\to K^{*+}_{2}(\to K^{0}\pi^+)\ell^+\ell^-$          &       $0.99\pm 0.45$                      &           $0.64\pm 0.13$                 &           $0.27\pm 0.10$  \\
$B^+_u\to K^{*+}_{2}(\to K^+\pi^0)\ell^+\ell^-$            &       $0.50\pm 0.23$                      &           $0.32\pm 0.06$                 &           $0.14\pm 0.05$  \\
$B^+_u\to K^{*+}_{2}(\to K^+\eta)\ell^+\ell^-$             &       $(9.93\pm 8.96)\times10^{-3}$       &           $(5.78\pm 4.90)\times10^{-3}$  &           $(2.92\pm 2.63)\times10^{-5}$  \\ \hline
$B^0_d\to K^{*0}_{2}(\to K^{+}\pi^-)\ell^+\ell^-$          &       $0.92\pm 0.43$                      &           $0.60\pm 0.11$                 &           $0.24\pm 0.09$  \\
$B^0_d\to K^{*0}_{2}(\to K^0\pi^0)\ell^+\ell^-$            &       $0.46\pm 0.21$                      &           $0.30\pm 0.06$                 &           $0.12\pm 0.05$  \\
$B^0_d\to K^{*0}_{2}(\to K^0\eta)\ell^+\ell^-$             &       $(9.19\pm 8.22)\times10^{-3}$       &           $(5.40\pm 4.59)\times10^{-3}$  &           $(2.57\pm 2.31)\times10^{-3}$  \\\hline
$B^0_s\to f_{2}(\to \pi^+\pi^-)\ell^+\ell^-$               &       $0.074\pm 0.038$                    &           $0.054\pm 0.020$               &           $0.13\pm 0.07$  \\
$B^0_s\to f_{2}(\to \pi^0\pi^0)\ell^+\ell^-$               &       $0.037\pm 0.019$                    &           $0.027\pm 0.010$               &           $0.066\pm 0.036$  \\
$B^0_s\to f_{2}(\to K^+K^-)\ell^+\ell^-$                   &       $(3.17\pm 1.79)\times10^{-3}$       &           $(2.31\pm 0.95)\times10^{-3}$  &           $(5.59\pm 3.21)\times10^{-3}$  \\
$B^0_s\to f_{2}(\to K^0\overline{K}^0)\ell^+\ell^-$        &       $(2.98\pm 1.69)\times10^{-3}$       &           $(2.18\pm 0.90)\times10^{-3}$  &           $(5.26\pm 3.02)\times10^{-3}$  \\
$B^0_s\to f_{2}(\to \eta\eta)\ell^+\ell^-$                 &       $(5.55\pm 3.42)\times10^{-4}$       &           $(3.99\pm 1.96)\times10^{-4}$  &           $(9.43\pm 5.84)\times10^{-4}$  \\\hline
$B^0_s\to f'_{2}(\to \pi^+\pi^-)\ell^+\ell^-$           &       $0.015\pm0.008$                     &           $(9.18\pm 2.89)\times10^{-3}$  &           $(3.99\pm 1.97)\times10^{-3}$  \\
$B^0_s\to f'_{2}(\to \pi^0\pi^0)\ell^+\ell^-$           &       $(7.49\pm 3.79)\times10^{-3}$       &           $(4.62\pm 1.45)\times10^{-3}$  &           $(2.00\pm 0.99)\times10^{-3}$  \\
$B^0_s\to f'_{2}(\to K^+K^-)\ell^+\ell^-$               &       $1.15\pm0.43$                       &           $0.72\pm0.11$                  &           $0.31\pm0.13$  \\
$B^0_s\to f'_{2}(\to K^0\overline{K}^0)\ell^+\ell^-$    &       $1.12\pm 0.42$                      &           $0.70\pm 0.11$                 &           $0.30\pm0.13$  \\
$B^0_s\to f'_{2}(\to \eta\eta)\ell^+\ell^-$             &       $0.31\pm 0.15$                      &           $0.19\pm 0.06$                 &           $0.084\pm 0.04$  \\
$B^0_s\to f'_{2}(\to \eta\eta')\ell^+\ell^-$            &       $(3.10\pm 3.09)\times10^{-3}$       &           $(1.96\pm 1.95)\times10^{-3}$  &           $(7.63\pm 7.57)\times10^{-4}$  \\\hline
\end{tabular}
\end{center}}\label{Tab:data1PPllTbs}
%\end{table}
%\begin{table}[htb]
\caption{Branching ratios for the $B\to PP\ell^+\ell^-$ decays with the tensor  resonances  under the $b\to d \ell^+\ell^-$ transitions within  1$\sigma$ error via the narrow width approximation. Ones for $\ell=e,\mu$ are  in unit of $10^{-7}$, and ones for $\ell=\tau$ are in unit of $10^{-9}$.}
\centering{\footnotesize
\begin{center}  \renewcommand\arraystretch{1.05}\tabcolsep0.3in
\begin{tabular}{lccc} \hline\hline
                                 ~~~~~~~~~~  Decay modes        &       $\mathcal{B}$ with  $\ell=e$        &      $\mathcal{B}$ with   $\ell=\mu$     &    $\mathcal{B}$  with $\ell=\tau$     \\ \hline
$B^+_u\to a^{+}_{2}(\to K^+\overline{K}^0)\ell^+\ell^-$         &       $(9.20\pm 4.64)\times10^{-3}$       &           $(6.30\pm 1.99)\times10^{-3}$  &           $(6.66\pm 3.42)\times10^{-3}$  \\
$B^+_u\to a^{+}_{2}(\to \pi^{+}\eta')\ell^+\ell^-$              &       $(9.76\pm 4.85)\times10^{-4}$       &           $(7.05\pm 2.29)\times10^{-4}$  &           $(7.50\pm 3.96)\times10^{-4}$  \\
$B^+_u\to a^{+}_{2}(\to \pi^{+}\eta)\ell^+\ell^-$               &       $0.026\pm0.012$                     &           $0.018\pm0.005$                &           $0.019\pm0.009$  \\\hline
$B^0_d\to a^{0}_{2}(\to K^+K^-)\ell^+\ell^-$                    &       $(2.17\pm 1.10)\times10^{-3}$       &           $(1.50\pm 0.47)\times10^{-3}$  &           $(1.58\pm 0.81)\times10^{-3}$  \\
$B^0_d\to a^{0}_{2}(\to K^0\overline{K}^0)\ell^+\ell^-$         &       $(2.06\pm 1.04)\times10^{-3}$       &           $(1.42\pm 0.44)\times10^{-3}$  &           $(1.50\pm 0.77)\times10^{-3}$  \\
$B^0_d\to a^{0}_{2}(\to \pi^0\eta')\ell^+\ell^-$                &       $(4.70\pm 2.32)\times10^{-4}$       &           $(3.33\pm 1.09)\times10^{-4}$  &           $(3.56\pm 1.87)\times10^{-4}$  \\
$B^0_d\to a^{0}_{2}(\to \pi^0\eta)\ell^+\ell^-$                 &       $0.012\pm 0.006$                    &           $(8.53\pm 2.21)\times10^{-3}$  &           $(8.94\pm 0.39)\times10^{-3}$  \\ \hline
$B^0_d\to f_{2}(\to \pi^+\pi^-)\ell^+\ell^-$                    &       $(1.38\pm 0.77)\times10^{-3}$       &           $(9.85\pm 3.70)\times10^{-4}$  &           $(1.36\pm 0.73)\times10^{-3}$  \\
$B^0_d\to f_{2}(\to \pi^0\pi^0)\ell^+\ell^-$                    &       $(6.94\pm 3.87)\times10^{-4}$       &           $(4.97\pm 1.87)\times10^{-4}$  &           $(6.87\pm 3.66)\times10^{-4}$  \\
$B^0_d\to f_{2}(\to K^+K^-)\ell^+\ell^-$                        &       $(5.99\pm 3.63)\times10^{-5}$       &           $(4.11\pm 1.70)\times10^{-5}$  &           $(5.78\pm 3.25)\times10^{-5}$  \\
$B^0_d\to f_{2}(\to K^0\overline{K}^0)\ell^+\ell^-$             &       $(5.64\pm 3.42)\times10^{-5}$       &           $(3.86\pm 1.60)\times10^{-5}$  &           $(5.44\pm 3.06)\times10^{-5}$  \\
$B^0_d\to f_{2}(\to \eta\eta)\ell^+\ell^-$                      &       $(1.03\pm 0.65)\times10^{-5}$       &           $(7.32\pm 3.67)\times10^{-6}$  &           $(1.01\pm 0.62)\times10^{-5}$  \\\hline
$B^0_d\to f^{'}_{2}(\to \pi^+\pi^-)\ell^+\ell^-$                &       $(2.77\pm 1.61)\times10^{-4}$       &           $(1.69\pm 0.57)\times10^{-4}$  &           $(2.52\pm 1.29)\times10^{-5}$  \\
$B^0_d\to f^{'}_{2}(\to \pi^0\pi^0)\ell^+\ell^-$                &       $(1.39\pm 0.81)\times10^{-4}$       &           $(8.51\pm 2.87)\times10^{-5}$  &           $(1.27\pm 0.65)\times10^{-5}$  \\
$B^0_d\to f^{'}_{2}(\to K^+K^-)\ell^+\ell^-$                    &       $0.021\pm 0.010$                    &           $0.013\pm 0.003$               &           $(1.90\pm 0.82)\times10^{-3}$  \\
$B^0_d\to f^{'}_{2}(\to K^0\overline{K}^0)\ell^+\ell^-$         &       $0.020\pm 0.010$                    &           $0.013\pm 0.003$               &           $(1.84\pm 0.79)\times10^{-3}$  \\
$B^0_d\to f^{'}_{2}(\to \eta\eta)\ell^+\ell^-$                  &       $(5.73\pm 3.24)\times10^{-3}$       &           $(3.51\pm 1.16)\times10^{-3}$  &           $(5.38\pm 2.80)\times10^{-4}$  \\
$B^0_d\to f^{'}_{2}(\to \eta\eta')\ell^+\ell^-$                 &       $(6.05\pm 6.02)\times10^{-5}$       &           $(3.71\pm 3.69)\times10^{-5}$  &           $(4.73\pm 4.69)\times10^{-6}$  \\\hline
$B^0_s\to \overline{K}^{*0}_{2}(\to K^{-}\pi^+)\ell^+\ell^-$              & $0.048\pm 0.023$                &           $0.033\pm0.007$                &           $0.029\pm 0.012$  \\
$B^0_s\to \overline{K}^{*0}_{2}(\to \overline{K}^0\pi^0)\ell^+\ell^-$     & $0.024\pm 0.011$                &           $0.016\pm0.003$                &           $0.015\pm 0.006$  \\
$B^0_s\to \overline{K}^{*0}_{2}(\to \overline{K}^0\eta)\ell^+\ell^-$      & $(4.89\pm 4.43)\times10^{-4}$   &           $(2.94\pm 2.52)\times10^{-4}$  &           $(3.08\pm 2.78)\times10^{-4}$  \\\hline
\end{tabular}
\end{center}}\label{Tab:data1PPllTbd}
\end{table}

\begin{table}[htb]
\caption{Branching ratios for the $B\to PV\ell^+\ell^-$ decays with the tensor  resonances  under the $b\to s \ell^+\ell^-$ transitions within  1$\sigma$ error via the narrow width approximation. Ones for $\ell=e,\mu$ are  in unit of $10^{-7}$, and ones for $\ell=\tau$ are in unit of $10^{-9}$. }
\centering{\footnotesize
\begin{center}  \renewcommand\arraystretch{1.05}\tabcolsep0.25in
\begin{tabular}{lccc}\hline\hline
                                 ~~~~~~~~~~  Decay modes   &       $\mathcal{B}$ with  $\ell=e$        &      $\mathcal{B}$ with   $\ell=\mu$     &    $\mathcal{B}$  with $\ell=\tau$     \\ \hline
 $B^+_u\to K^{*+}_{2}(\to K^{*+}\pi^0)\ell^+\ell^-$              &     $0.25\pm 0.12$       &     $0.16\pm 0.04$     &     $0.070\pm 0.029$  \\
 $B^+_u\to K^{*+}_{2}(\to K^{*0}\pi^+)\ell^+\ell^-$              &     $0.48\pm 0.23$       &     $0.31\pm 0.07$     &     $0.14\pm 0.06$  \\
 $B^+_u\to K^{*+}_{2}(\to \rho^0K^+)\ell^+\ell^-$                &     $0.087\pm 0.043$     &     $0.057\pm 0.015$   &     $0.024\pm 0.010$  \\
 $B^+_u\to K^{*+}_{2}(\to \rho^+K^0)\ell^+\ell^-$                &     $0.17\pm 0.08$       &     $0.11\pm 0.03$     &     $0.047\pm 0.019$  \\
 $B^+_u\to K^{*+}_{2}(\to \omega K^+)\ell^+\ell^-$               &     $0.092\pm 0.055$     &     $0.058\pm 0.024$   &     $0.025\pm 0.013$  \\
 $B^+_u\to K^{*+}_{2}(\to K^{*+}\eta)\ell^+\ell^-$               &     $0.017\pm 0.009$     &     $0.011\pm 0.003$   &     $(5.01\pm 2.32)\times10^{-3}$  \\
 $B^+_u\to K^{*+}_{2}(\to \phi K^{+})\ell^+\ell^-$               &     $(1.16\pm 0.65)\times10^{-3}$     &     $(7.61\pm 2.52)\times10^{-4}$     &     $(3.33\pm 1.62)\times10^{-4}$  \\   \hline
 $B^0_d\to K^{*0}_{2}(\to K^{*+}\pi^-)\ell^+\ell^-$              &     $0.47\pm 0.23$       &     $0.31\pm 0.07$     &     $0.13\pm 0.05$  \\
 $B^0_d\to K^{*0}_{2}(\to K^{*0}\pi^0)\ell^+\ell^-$              &     $0.23\pm 0.11$       &     $0.15\pm 0.03$     &     $0.062\pm 0.026$  \\
 $B^0_d\to K^{*0}_{2}(\to \rho^-K^{+})\ell^+\ell^-$              &     $0.17\pm 0.08$       &     $0.11\pm 0.03$     &     $0.045\pm 0.019$  \\
 $B^0_d\to K^{*0}_{2}(\to \rho^0K^{0})\ell^+\ell^-$              &     $0.080\pm 0.040$     &     $0.053\pm 0.013$     &     $0.022\pm 0.009$  \\
 $B^0_d\to K^{*0}_{2}(\to \omega K^{0})\ell^+\ell^-$             &     $0.085\pm 0.052$     &     $0.055\pm 0.022$     &     $0.023\pm 0.012$  \\
 $B^0_d\to K^{*0}_{2}(\to K^{*0}\eta)\ell^+\ell^-$               &     $0.016\pm 0.009$     &     $0.010\pm 0.003$     &     $(4.33\pm 2.06)\times10^{-3}$  \\
 $B^0_d\to K^{*0}_{2}(\to \phi K^0)\ell^+\ell^-$                 &     $(1.17\pm 0.66)\times10^{-3}$     &     $(7.25\pm 2.37)\times10^{-4}$     &     $(3.08\pm 1.51)\times10^{-4}$  \\  \hline
 $B^0_s\to f_{2}(\to K^{*+}K^-/K^{*-}K^+)\ell^+\ell^-$            &     $(9.36\pm 5.00)\times10^{-5}$     &     $(6.91\pm 2.57)\times10^{-5}$     &     $(1.62\pm 0.87)\times10^{-4}$  \\
 $B^0_s\to f_{2}(\to K^{*0}\overline{K}^0/\overline{K}^{*0}K^0)\ell^+\ell^-$          &     $(8.29\pm 4.46)\times10^{-5}$     &     $(6.13\pm 2.31)\times10^{-5}$     &     $(1.43\pm 0.77)\times10^{-4}$  \\
 $B^0_s\to f_{2}(\to K^*K)\ell^+\ell^-$                           &     $(3.53\pm 1.89)\times10^{-4}$     &     $(2.61\pm 0.98)\times10^{-4}$     &     $(6.09\pm 3.28)\times10^{-4}$  \\  \hline
 $B^0_s\to f^{'}_{2}(\to K^{*+}K^-/K^{*-}K^+)\ell^+\ell^-$      &     $0.17\pm 0.08$       &     $0.11\pm 0.04$     &     $0.045\pm 0.020$  \\
 $B^0_s\to f^{'}_{2}(\to K^{*0}\overline{K}^0/\overline{K}^{*0}K^0)\ell^+\ell^-$      &     $0.15\pm 0.07$       &     $0.096\pm 0.032$     &     $0.040\pm 0.017$  \\
 $B^0_s\to f^{'}_{2}(\to K^*K)\ell^+\ell^-$                       &     $0.63\pm 0.31$       &     $0.41\pm 0.14$     &     $0.17\pm 0.07$  \\\hline
\end{tabular}
\end{center}}\label{Tab:data1PVllTbs}
%\end{table}
%\begin{table}[htb]
\caption{Branching ratios for the $B\to PV\ell^+\ell^-$ decays with the tensor  resonances  under the $b\to d \ell^+\ell^-$ transitions within  1$\sigma$ error via the narrow width approximation. Ones for $\ell=e,\mu$ are  in unit of $10^{-7}$, and ones for $\ell=\tau$ are in unit of $10^{-9}$.}
\centering{\footnotesize
\begin{center}  \renewcommand\arraystretch{1.05}\tabcolsep0.25in
\begin{tabular}{lccc}\hline\hline
                                  ~~~~~~~~~~  Decay modes   &       $\mathcal{B}$ with  $\ell=e$        &      $\mathcal{B}$ with   $\ell=\mu$     &    $\mathcal{B}$  with $\ell=\tau$     \\ \hline
 $B^+_u\to a^{+}_{2}(\to \rho^{0}\pi^+)\ell^+\ell^-$                &     $0.064\pm 0.028$                  &     $0.044\pm 0.010$                  &     $0.046\pm 0.019$  \\
 $B^+_u\to a^{+}_{2}(\to \rho^{+}\pi^0)\ell^+\ell^-$                &     $0.064\pm 0.028$                  &     $0.045\pm 0.010$                  &     $0.046\pm 0.019$  \\
 $B^+_u\to a^{+}_{2}(\to \overline{K}^{*0}K^+)\ell^+\ell^-$         &     $(1.95\pm 1.04)\times10^{-4}$     &     $(1.32\pm 0.46)\times10^{-4}$     &     $(1.43\pm 0.71)\times10^{-4}$  \\
 $B^+_u\to a^{+}_{2}(\to K^{*+}\overline{K}^0)\ell^+\ell^-$         &     $(2.04\pm 1.07)\times10^{-4}$     &     $(1.41\pm 0.48)\times10^{-4}$     &     $(1.50\pm 0.74)\times10^{-4}$  \\  \hline
 $B^0_d\to a^{0}_{2}(\to \rho^-\pi^+)\ell^+\ell^-$                  &     $0.030\pm 0.013$                  &     $0.020\pm 0.004$                  &     $0.021\pm 0.009$  \\
 $B^0_d\to a^{0}_{2}(\to \rho^+\pi^-)\ell^+\ell^-$                  &     $0.030\pm 0.013$                  &     $0.020\pm 0.004$                  &     $0.021\pm 0.009$  \\
 $B^0_d\to a^{0}_{2}(\to K^{*+}K^-/K^{*-}K^+)\ell^+\ell^-$          &     $(5.11\pm 2.72)\times10^{-5}$     &     $(3.53\pm 1.20)\times10^{-5}$     &     $(3.78\pm 1.85)\times10^{-5}$  \\
 $B^0_d\to a^{0}_{2}(\to K^{*0}\overline{K}^0/\overline{K}^{*0}K^0)\ell^+\ell^-$        &     $(4.09\pm 2.25)\times10^{-5}$     &     $(2.79\pm 0.99)\times10^{-5}$     &     $(3.02\pm 1.51)\times10^{-5}$  \\\hline
 $B^0_d\to f_{2}(\to K^{*+}K^-/K^{*-}K^+)\ell^+\ell^-$              &   $(1.77\pm 1.02)\times10^{-6}$       &   $(1.24\pm 0.47)\times10^{-6}$       &   $(1.73\pm 0.94)\times10^{-6}$               \\
 $B^0_d\to f_{2}(\to K^{*0}\overline{K}^0/\overline{K}^{*0}K^0)\ell^+\ell^-$            &   $(1.57\pm 0.91)\times10^{-6}$       &   $(1.09\pm 0.42)\times10^{-6}$       &   $(1.54\pm 0.84)\times10^{-6}$               \\
  $B^0_d\to f_{2}(\to K^{*}K)\ell^+\ell^-$                          &   $(6.69\pm 3.86)\times10^{-6}$       &   $(4.66\pm 1.78)\times10^{-6}$       &   $(6.54\pm 3.55)\times10^{-6}$               \\  \hline
 $B^0_d\to f'_{2}(\to K^{*+}K^-/K^{*-}K^+)\ell^+\ell^-$             &   $(3.19\pm 1.80)\times10^{-3}$       &   $(1.96\pm 0.68)\times10^{-3}$       &   $(2.73\pm 1.19)\times10^{-4}$             \\
 $B^0_d\to f'_{2}(\to K^{*0}\overline{K}^0/\overline{K}^{*0}K^0)\ell^+\ell^-$           &   $(2.84\pm 1.61)\times10^{-3}$       &   $(1.73\pm 0.60)\times10^{-3}$       &   $(2.43\pm 1.05)\times10^{-4}$             \\
  $B^0_d\to f'_{2}(\to K^{*}K)\ell^+\ell^-$                         &   $0.012\pm0.007$                     &   $(7.36\pm 2.56)\times10^{-3}$       &   $(1.03\pm 0.45)\times10^{-3}$                           \\  \hline
 $B^0_s\to \overline{K}^{*0}_{2}(\to K^{*-}\pi^+)\ell^+\ell^-$              &     $0.025\pm 0.012$                  &     $0.017\pm 0.004$                  &     $0.015\pm 0.006$  \\
 $B^0_s\to \overline{K}^{*0}_{2}(\to \overline{K}^{*0}\pi^0)\ell^+\ell^-$   &     $0.012\pm 0.006$                  &     $(8.23\pm 1.89)\times10^{-3}$     &     $(7.33\pm 3.18)\times10^{-3}$  \\
 $B^0_s\to \overline{K}^{*0}_{2}(\to \rho^+K^-)\ell^+\ell^-$                &     $(8.72\pm 4.26)\times10^{-3}$     &     $(5.98\pm 1.54)\times10^{-3}$     &     $(5.42\pm 2.33)\times10^{-3}$  \\
 $B^0_s\to \overline{K}^{*0}_{2}(\to \rho^0\overline{K}^0)\ell^+\ell^-$     &     $(4.23\pm 2.08)\times10^{-3}$     &     $(2.89\pm 0.75)\times10^{-3}$     &     $(2.60\pm 1.12)\times10^{-3}$  \\
 $B^0_s\to \overline{K}^{*0}_{2}(\to \omega\overline{K}^0)\ell^+\ell^-$     &     $(4.57\pm 2.70)\times10^{-3}$     &     $(2.96\pm 1.25)\times10^{-3}$     &     $(2.67\pm 1.48)\times10^{-3}$  \\
 $B^0_s\to \overline{K}^{*0}_{2}(\to \overline{K}^{*0}\eta)\ell^+\ell^-$    &     $(8.29\pm 4.40)\times10^{-4}$     &     $(5.54\pm 1.60)\times10^{-4}$     &     $(5.20\pm 2.46)\times10^{-4}$  \\
 $B^0_s\to \overline{K}^{*0}_{2}(\to \phi \overline{K}^{0})\ell^+\ell^-$    &     $(6.02\pm 3.34)\times10^{-5}$     &     $(3.93\pm 1.32)\times10^{-5}$     &     $(3.67\pm 1.82)\times10^{-5}$  \\ \hline 		
\end{tabular}
\end{center}}\label{Tab:data1PVllTbd}
\end{table}

%\newpage

For the $B\to T(\to PP)\ell^+\ell^-$ decays with $\ell=e,\mu$, only $\mathcal{B}(B^0_s\to f'_2(\to K^0\overline{K}^0)e^+e^-)$ and $\mathcal{B}(B^0_s\to f'_2(\to K^+K^-)e^+e^-)$ are on the order of $\mathcal{O}(10^{-7})$, the others are on the order of $10^{-8}$ or smaller.  Our prediction  $\mathcal{B}(B^0_s\to f'_2(\to \pi^+\pi^-)\mu^+\mu^-)=(9.18\pm2.89)\times10^{-10}$,  which is much smaller than corresponding total branching ratios   $\mathcal{B}(B^0_s\to  \pi^+\pi^-\mu^+\mu^-)=(8.4\pm1.7)\times10^{-8}$ from PDG \cite{PDG2025} and  $\mathcal{B}(B^0_s\to f_0(980)(\to \pi^+\pi^-)\mu^+\mu^-)=(8.3\pm1.7)\times10^{-8}$ from  LHCb \cite{LHCb:2014yov}.  Similarly, our prediction  $\mathcal{B}(B^0_d\to f_2(\to \pi^+\pi^-)\mu^+\mu^-)=(9.85\pm3.70)\times10^{-11}$ and $\mathcal{B}(B^0_d\to f'_2(\to \pi^+\pi^-)\mu^+\mu^-)=(1.69\pm0.57)\times10^{-11}$  are also  much smaller than   $\mathcal{B}(B^0_d\to  \pi^+\pi^-\mu^+\mu^-)=(2.1\pm0.5)\times10^{-8}$ from PDG \cite{PDG2025}.

For the $B\to T(\to PV)\ell^+\ell^-$ decays, their branching ratio estimates  are on the order of $10^{-13}-10^{-8}$ for $\ell=e,\mu$ and $\mathcal{O}(10^{-16})-\mathcal{O}(10^{-10})$ for $\ell=\tau$. Note that the $B^+_u\to  \phi K^+\mu^+\mu^-$ decay has been measured \cite{PDG2025},  $\mathcal{B}(B^+_u\to  \phi K^+\mu^+\mu^-)=(7.9^{+2.1}_{-1.7})\times10^{-8}$, which is much larger than our prediction $\mathcal{B}(B^+_u\to K^{*+}_2(\to \phi K^+)\mu^+\mu^-)=(7.61\pm2.52)\times10^{-11}$.

The comparison with the available total branching fractions indicates that the tensor resonance contributions are generally subdominant in the measured $B\to PP\ell^+\ell^-$ and $B\to PV\ell^+\ell^-$ modes. For example, the tensor contribution to $B_{s}^0\to \pi^+\pi^-\mu^+\mu^-$ is much smaller than the measured total rate, suggesting that scalar or vector resonances may dominate this channel. Similarly, the tensor contribution to $B_u^+\to \phi K^+\mu^+\mu^-$ is far below the measured total branching fraction, and other resonances, such as axial-vector states or their excitations, are expected to provide the leading contribution. Nevertheless, the smallness of the tensor contribution does not mean that it is experimentally inaccessible.
Since tensor resonances carry $J^P=2^+$, their contributions can in principle be distinguished from  scalar, vector, and axial-vector components  by combining invariant mass distributions with angular or partial-wave analyses \cite{LHCb:2016eyu,Blake:2012mb}. In this sense, the tensor resonance estimates obtained here should be regarded as inputs for future partial wave or amplitude analyses, rather than as predictions for the full four body decay ratios.

On the other hand, the $B^0_s\to K\bar K\ell^+\ell^-$ decays  receive relatively larger tensor resonance contributions in our estimates. These modes may therefore be more promising for testing the tensor components,
especially through invariant mass and angular analyses. In such analyses, observables sensitive to the spin-2 angular structure, such as angular moments or partial-wave fractions, would be more useful than the total branching fraction alone.

After the conpletion of this work, the LHCb reported   searches for $B^0\to K^+\pi^-\tau^+\tau^-$, $B^0_s\to K^+K^-\tau^+\tau^-$ and $B^0_s\to K^+\pi^-\tau^+\tau^-$  decays \cite{LHCb:2025lcw}, the upper limits of their total branching ratios with different bins are reported, and they are at the order of $10^{-6}-10^{-4}$. Our relevant  estimates with the tensor resonances are much smaller than their total experimental upper limits.

\subsection{ $\mathcal{B}(B\to PP/PV\ell^+\ell^-)$ with the  width effects }
 Some decay widths of the tensor mesons and vector mesons are not very narrow.  We will take into account  the width effects of both the resonance  states and   vector mesons in the final state of the $T\to PV$ decay process.
The decay branching ratios of $B\to M_1M_2\ell^+\ell^-$ with  the width effects of the resonance  states can be written as  \cite{Cheng:1993ah,Tsai:2021ota}
{\small
\begin{eqnarray}
\mathcal{B}(B\to R(\to M_1M_2)\ell^+\ell^-)=\frac{1}{\pi}\int_{(m_{R}-n\Gamma_{R})^2}^{(m_{R}+n\Gamma_{R})^2} dt_R \int_{4m^2_{\ell}}^{(m_B-\sqrt{t_R})^2} dq^2\frac{\sqrt{t_R}d\mathcal{B}(B\to R\ell^+\ell^-,t_R)/dq^2~\mathcal{B}(R\to M_1M_2,t_R)\Gamma_{R}}{(t_R-m_{R}^2)^2+m^2_{R}\Gamma^2_{R}},\label{Eq:RDA4}
\end{eqnarray}}
where $R$ denotes the resonance  states of the  tensor mesons, $d\mathcal{B}(B\to R\ell^+\ell^-,t_R)/dq^2$ and $\mathcal{B}(R\to M_1M_2,t_R)$ are obtained from  $d\mathcal{B}(B\to R\ell^+\ell^-)/dq^2$ in Eq. (\ref{Eq:dBr}) and $\mathcal{B}(R\to M_1M_2)$ in Refs. \cite{Qiao:2024nbq,RMWB2PPlv} by replacing $m_R \to \sqrt{t_R}$, respectively. We choose $n=2$, and the width effects of vector mesons are considered in accordance with the framework established in Refs. \cite{Qiao:2024nbq,RMWB2PPlv}.
In addition, in the finite width calculation, the threshold function $\Theta \left[t_R-(m_1+m_2)^2\right]$ has been added in Eq. (\ref{Eq:RDA4}) to  ensure that each channel contributes only above
its physical threshold.

\begin{table}[htb]
\caption{Branching ratios for the $B\to PP\ell^+\ell^-$ decays with the  finite width effects under the $b\to s \ell^+\ell^-$ transitions  within  1$\sigma$ error. Ones for $\ell=e,\mu$ are  in unit of $10^{-7}$, and ones for $\ell=\tau$ are in unit of $10^{-9}$. $^\sharp$denotes that the results should be regards as qualitative estimates since we used the  fixed total width.}
\centering{\footnotesize
\begin{center}  \renewcommand\arraystretch{1.05}\tabcolsep0.3in
\begin{tabular}{lccc} \hline\hline
~~~~~~~~~~  Decay modes   &       $\mathcal{B}$ with  $\ell=e$        &      $\mathcal{B}$ with   $\ell=\mu$     &    $\mathcal{B}$  with $\ell=\tau$     \\ \hline
$B^+_u\to K^{*+}_{2}(\to K^{0}\pi^+)\ell^+\ell^-$          &       $0.84\pm 0.32$                         &           $0.54\pm 0.09$                         &           $0.26\pm0.11$  \\
$B^+_u\to K^{*+}_{2}(\to K^+\pi^0)\ell^+\ell^-$            &       $0.43\pm 0.16$                         &           $0.28\pm 0.05$                         &           $0.13\pm0.05$  \\
$B^+_u\to K^{*+}_{2}(\to K^+\eta)\ell^+\ell^-$             &       $(8.02\pm 6.99)\times10^{-3}$          &           $(4.94\pm 4.19)\times10^{-3}$          &           $(2.16\pm 1.92)\times10^{-3}$  \\\hline
$B^0_d\to K^{*0}_{2}(\to K^{+}\pi^-)\ell^+\ell^-$          &       $0.79\pm 0.30$                         &           $0.51\pm 0.09$                         &           $0.24\pm0.10$  \\
$B^0_d\to K^{*0}_{2}(\to K^0\pi^0)\ell^+\ell^-$            &       $0.39\pm 0.15$                         &           $0.25\pm 0.04$                         &           $0.12\pm0.05$  \\
$B^0_d\to K^{*0}_{2}(\to K^0\eta)\ell^+\ell^-$             &       $(7.38\pm 6.43)\times10^{-3}$          &           $(4.55\pm 3.86)\times10^{-3}$          &           $(1.88\pm 1.67)\times10^{-3}$  \\ \hline
$B^0_s\to f_{2}(\to \pi^+\pi^-)\ell^+\ell^-$               &       $0.061\pm 0.025$                       &           $0.045\pm 0.016$                       &           $0.12\pm0.06$  \\
$B^0_s\to f_{2}(\to \pi^0\pi^0)\ell^+\ell^-$               &       $0.031\pm 0.012$                       &           $0.023\pm 0.008$                       &           $0.060\pm0.029$  \\
$B^0_s\to f_{2}(\to K^+K^-)\ell^+\ell^-$                   &       $(2.79\pm 1.27)\times10^{-3}$          &           $(2.11\pm 0.90)\times10^{-3}$          &           $(3.54\pm 1.83)\times10^{-3}$  \\
$B^0_s\to f_{2}(\to K^0\overline{K}^0)\ell^+\ell^-$        &       $(2.65\pm 1.21)\times10^{-3}$          &           $(2.00\pm 0.86)\times10^{-3}$          &           $(3.32\pm 1.72)\times10^{-3}$  \\
$B^0_s\to f_{2}(\to \eta\eta)\ell^+\ell^-$                 &       $(6.20\pm 3.33)\times10^{-4}$          &           $(4.43\pm 2.05)\times10^{-4}$          &           $(6.11\pm 3.43)\times10^{-4}$  \\\hline
$B^0_s\to f'_{2}(\to \pi^+\pi^-)\ell^+\ell^-$           &       $0.013\pm 0.006$                       &           $(7.80\pm 2.32)\times10^{-3}$          &           $(3.84\pm 1.96)\times10^{-3}$  \\
$B^0_s\to f'_{2}(\to \pi^0\pi^0)\ell^+\ell^-$           &       $(6.40\pm 2.86)\times10^{-3}$          &           $(3.92\pm 1.16)\times10^{-3}$          &           $(1.93\pm 0.99)\times10^{-3}$  \\
$B^0_s\to f'_{2}(\to K^+K^-)\ell^+\ell^-$               &       $0.95\pm 0.36$                         &           $0.61\pm 0.10$                         &           $0.27\pm0.12$  \\
$B^0_s\to f'_{2}(\to K^0\overline{K}^0)\ell^+\ell^-$    &       $0.92\pm 0.35$                         &           $0.59\pm 0.09$                         &           $0.27\pm0.12$  \\
$B^0_s\to f'_{2}(\to \eta\eta)\ell^+\ell^-$             &       $0.26\pm 0.13$                         &           $0.16\pm 0.05$                         &           $0.076\pm0.042$  \\
$B^0_s\to f'_{2}(\to \eta\eta')\ell^+\ell^-$            &       $0.039\pm 0.036$$^\sharp$                       &           $0.023\pm 0.021$$^\sharp$                       &           $(4.05\pm 3.68)\times10^{-3}$$^\sharp$  \\\hline
\end{tabular}
\end{center}}\label{Tab:data1PPllTbsWE}
%\end{table}
%\begin{table}[htb]
\caption{Branching ratios for the $B\to PP\ell^+\ell^-$ decays with the  finite width effects   under the $b\to d \ell^+\ell^-$ transitions within  1$\sigma$ error. Ones for $\ell=e,\mu$ are  in unit of $10^{-7}$, and ones for $\ell=\tau$ are in unit of $10^{-9}$. $^\sharp$denotes that the results should be regards as qualitative estimates since we used the  fixed total width.}
\centering{\footnotesize
\begin{center}  \renewcommand\arraystretch{1.05}\tabcolsep0.3in
\begin{tabular}{lccc} \hline\hline
~~~~~~~~~~  Decay modes        &       $\mathcal{B}$ with  $\ell=e$        &      $\mathcal{B}$ with   $\ell=\mu$     &    $\mathcal{B}$  with $\ell=\tau$     \\ \hline
$B^+_u\to a^{+}_{2}(\to K^+\overline{K}^0)\ell^+\ell^-$         &       $(8.00\pm 3.40)\times10^{-3}$          &           $(5.32\pm 1.66)\times10^{-3}$          &           $(5.26\pm 2.59)\times10^{-3}$  \\
$B^+_u\to a^{+}_{2}(\to \pi^{+}\eta')\ell^+\ell^-$              &       $(1.04\pm 0.45)\times10^{-3}$          &           $(6.80\pm 2.00)\times10^{-4}$          &           $(5.67\pm 2.72)\times10^{-4}$  \\
$B^+_u\to a^{+}_{2}(\to \pi^{+}\eta)\ell^+\ell^-$               &       $0.023\pm 0.008$                       &           $0.015\pm 0.004$                       &           $0.016\pm0.006$  \\\hline
$B^0_d\to a^{0}_{2}(\to K^+K^-)\ell^+\ell^-$                    &       $(1.90\pm 0.81)\times10^{-3}$          &           $(1.26\pm 0.39)\times10^{-3}$          &           $(1.25\pm 0.62)\times10^{-3}$  \\
$B^0_d\to a^{0}_{2}(\to K^0\overline{K}^0)\ell^+\ell^-$         &       $(1.81\pm 0.77)\times10^{-3}$          &           $(1.20\pm 0.37)\times10^{-3}$          &           $(1.18\pm 0.58)\times10^{-3}$  \\
$B^0_d\to a^{0}_{2}(\to \pi^0\eta')\ell^+\ell^-$                &       $(4.93\pm 2.13)\times10^{-4}$          &           $(3.23\pm 0.95)\times10^{-4}$          &           $(2.71\pm 1.29)\times10^{-4}$  \\
$B^0_d\to a^{0}_{2}(\to \pi^0\eta)\ell^+\ell^-$                 &       $0.011\pm 0.004$                       &           $(7.08\pm 1.71)\times10^{-3}$          &           $(7.45\pm 3.02)\times10^{-3}$ \\\hline
$B^0_d\to f_{2}(\to \pi^+\pi^-)\ell^+\ell^-$                    &       $(1.14\pm 0.49)\times10^{-3}$          &           $(8.13\pm 2.97)\times10^{-4}$          &           $(1.31\pm 0.64)\times10^{-3}$  \\
$B^0_d\to f_{2}(\to \pi^0\pi^0)\ell^+\ell^-$                    &       $(5.76\pm 2.45)\times10^{-4}$          &           $(4.10\pm 1.50)\times10^{-4}$          &           $(6.62\pm 3.23)\times10^{-4}$  \\
$B^0_d\to f_{2}(\to K^+K^-)\ell^+\ell^-$                        &       $(5.29\pm 2.57)\times10^{-5}$          &           $(3.76\pm 1.57)\times10^{-5}$          &           $(3.60\pm 1.84)\times10^{-5}$  \\
$B^0_d\to f_{2}(\to K^0\overline{K}^0)\ell^+\ell^-$             &       $(5.03\pm 2.44)\times10^{-5}$          &           $(3.57\pm 1.49)\times10^{-5}$          &           $(3.37\pm 1.72)\times10^{-5}$  \\
$B^0_d\to f_{2}(\to \eta\eta)\ell^+\ell^-$                      &       $(1.20\pm 0.67)\times10^{-5}$          &           $(8.06\pm 3.85)\times10^{-6}$          &           $(5.86\pm 3.27)\times10^{-6}$ \\\hline
$B^0_d\to f^{'}_{2}(\to \pi^+\pi^-)\ell^+\ell^-$                &       $(2.42\pm 1.14)\times10^{-4}$          &           $(1.41\pm 0.44)\times10^{-4}$          &           $(2.57\pm 1.36)\times10^{-5}$  \\
$B^0_d\to f^{'}_{2}(\to \pi^0\pi^0)\ell^+\ell^-$                &       $(1.22\pm 0.57)\times10^{-4}$          &           $(7.11\pm 2.21)\times10^{-5}$          &           $(1.29\pm 0.68)\times10^{-5}$  \\
$B^0_d\to f^{'}_{2}(\to K^+K^-)\ell^+\ell^-$                    &       $0.018\pm 0.007$                       &           $0.011\pm 0.002$                       &           $(1.81\pm 0.80)\times10^{-3}$  \\
$B^0_d\to f^{'}_{2}(\to K^0\overline{K}^0)\ell^+\ell^-$         &       $0.018\pm 0.007$                       &           $0.011\pm 0.002$                       &           $(1.75\pm 0.77)\times10^{-3}$  \\
$B^0_d\to f^{'}_{2}(\to \eta\eta)\ell^+\ell^-$                  &       $(5.08\pm 2.65)\times10^{-3}$          &           $(2.93\pm 0.97)\times10^{-3}$          &           $(4.94\pm 2.80)\times10^{-4}$  \\
$B^0_d\to f^{'}_{2}(\to \eta\eta')\ell^+\ell^-$                 &       $(7.41\pm 6.86)\times10^{-4}$$^\sharp$          &           $(4.24\pm 3.81)\times10^{-4}$$^\sharp$          &           $(1.58\pm 1.44)\times10^{-5}$$^\sharp$ \\\hline
$B^0_s\to \overline{K}^{*0}_{2}(\to K^{-}\pi^+)\ell^+\ell^-$              &       $0.037\pm 0.012$                       &           $0.026\pm 0.005$                       &           $0.046\pm0.020$  \\
$B^0_s\to \overline{K}^{*0}_{2}(\to \overline{K}^0\pi^0)\ell^+\ell^-$     &       $0.019\pm 0.006$                       &           $0.013\pm 0.002$                       &           $0.023\pm0.010$  \\
$B^0_s\to \overline{K}^{*0}_{2}(\to \overline{K}^0\eta)\ell^+\ell^-$      &       $(2.54\pm 2.20)\times10^{-4}$          &           $(1.74\pm 1.48)\times10^{-4}$          &           $(2.72\pm2.43)\times10^{-4}$  \\ \hline
\end{tabular}
\end{center}}\label{Tab:data1PPllTbdWE}
\end{table}

\begin{table}[htb]
\caption{Branching ratios for the $B\to PV\ell^+\ell^-$ decays with the  finite width effects  under the $b\to s \ell^+\ell^-$ transitions within  1$\sigma$ error. Ones for $\ell=e,\mu$ are  in unit of $10^{-7}$, and ones for $\ell=\tau$ are in unit of $10^{-9}$. $^\sharp$denotes that the results should be regards as qualitative estimates since we used the  fixed total width. }
\centering{\footnotesize
\begin{center}  \renewcommand\arraystretch{1}\tabcolsep0.25in
\begin{tabular}{lccc}\hline\hline
	~~~~~~~~~~  Decay modes   &       $\mathcal{B}$ with  $\ell=e$        &      $\mathcal{B}$ with   $\ell=\mu$     &    $\mathcal{B}$  with $\ell=\tau$     \\ \hline
	$B^+_u\to K^{*+}_{2}(\to K^{*+}\pi^0)\ell^+\ell^-$              &     $0.21\pm 0.08$       &     $0.13\pm 0.03$     &     $0.050\pm 0.022$  \\
	$B^+_u\to K^{*+}_{2}(\to K^{*0}\pi^+)\ell^+\ell^-$              &     $0.39\pm 0.15$       &     $0.25\pm 0.05$     &     $0.096\pm 0.042$  \\
	$B^+_u\to K^{*+}_{2}(\to \rho^0K^+)\ell^+\ell^-$                &     $0.068\pm 0.028$     &     $0.043\pm 0.010$   &     $0.014\pm 0.006$  \\
	$B^+_u\to K^{*+}_{2}(\to \rho^+K^0)\ell^+\ell^-$                &     $0.13\pm 0.05$       &     $0.083\pm 0.019$     &     $0.028\pm 0.012$  \\
	$B^+_u\to K^{*+}_{2}(\to \omega K^+)\ell^+\ell^-$               &     $0.066\pm 0.033$     &     $0.042\pm 0.017$   &     $0.011\pm 0.006$  \\
	$B^+_u\to K^{*+}_{2}(\to K^{*+}\eta)\ell^+\ell^-$               &     $0.012\pm 0.005$$^\sharp$     &     $(6.85\pm 1.63)\times10^{-3}$$^\sharp$   &     $(8.79\pm 4.03)\times10^{-4}$$^\sharp$  \\
	$B^+_u\to K^{*+}_{2}(\to \phi K^{+})\ell^+\ell^-$               &     $(6.89\pm 3.11)\times10^{-4}$$^\sharp$      &     $(4.11\pm 1.14)\times10^{-4}$$^\sharp$      &     $(1.59\pm 0.61)\times10^{-5}$$^\sharp$   \\   \hline
	$B^0_d\to K^{*0}_{2}(\to K^{*+}\pi^-)\ell^+\ell^-$              &     $0.39\pm 0.15$       &     $0.24\pm 0.05$     &     $0.090\pm 0.039$  \\
	$B^0_d\to K^{*0}_{2}(\to K^{*0}\pi^0)\ell^+\ell^-$              &     $0.19\pm 0.07$       &     $0.12\pm 0.02$     &     $0.044\pm 0.019$  \\
	$B^0_d\to K^{*0}_{2}(\to \rho^-K^{+})\ell^+\ell^-$              &     $0.13\pm 0.05$       &     $0.083\pm 0.019$     &     $0.027\pm 0.012$  \\
	$B^0_d\to K^{*0}_{2}(\to \rho^0K^{0})\ell^+\ell^-$              &     $0.064\pm 0.026$     &     $0.040\pm 0.009$     &     $0.013\pm 0.005$  \\
	$B^0_d\to K^{*0}_{2}(\to \omega K^{0})\ell^+\ell^-$             &     $0.062\pm 0.031$     &     $0.040\pm 0.016$     &     $(9.81\pm 5.23)\times10^{-3}$  \\
	$B^0_d\to K^{*0}_{2}(\to K^{*0}\eta)\ell^+\ell^-$               &     $0.010\pm 0.004$$^\sharp$      &     $(6.16\pm 1.44)\times10^{-3}$$^\sharp$      &     $(7.27\pm 3.16)\times10^{-4}$$^\sharp$   \\
	$B^0_d\to K^{*0}_{2}(\to \phi K^0)\ell^+\ell^-$                 &     $(6.76\pm 3.20)\times10^{-4}$$^\sharp$      &     $(3.98\pm 1.08)\times10^{-4}$$^\sharp$      &     $(1.40\pm 0.55)\times10^{-5}$$^\sharp$   \\  \hline
	$B^0_s\to f_{2}(\to K^{*+}K^-/K^{*-}K^+)\ell^+\ell^-$            &     $(4.43\pm 2.11)\times10^{-5}$$^\sharp$      &     $(2.60\pm 0.82)\times10^{-5}$$^\sharp$      &     $(8.76\pm 3.78)\times10^{-6}$$^\sharp$   \\
	$B^0_s\to f_{2}(\to K^{*0}\overline{K}^0/\overline{K}^{*0}K^0)\ell^+\ell^-$          &     $(3.85\pm 1.84)\times10^{-5}$$^\sharp$      &     $(2.27\pm 0.73)\times10^{-5}$$^\sharp$      &     $(7.11\pm 3.08)\times10^{-6}$$^\sharp$   \\
	$B^0_s\to f_{2}(\to K^*K)\ell^+\ell^-$                           &     $(1.66\pm 0.79)\times10^{-4}$$^\sharp$      &     $(9.74\pm 3.11)\times10^{-5}$$^\sharp$      &     $(3.18\pm 1.37)\times10^{-5}$$^\sharp$   \\  \hline
	$B^0_s\to f^{'}_{2}(\to K^{*+}K^-/K^{*-}K^+)\ell^+\ell^-$      &     $0.13\pm 0.05$       &     $0.082\pm 0.024$     &     $0.026\pm 0.011$  \\
	$B^0_s\to f^{'}_{2}(\to K^{*0}\overline{K}^0/\overline{K}^{*0}K^0)\ell^+\ell^-$      &     $0.11\pm 0.05$       &     $0.072\pm 0.021$     &     $0.022\pm 0.010$  \\
	$B^0_s\to f^{'}_{2}(\to K^*K)\ell^+\ell^-$                       &     $0.47\pm 0.20$       &     $0.31\pm 0.09$     &     $0.096\pm 0.041$  \\\hline
\end{tabular}
\end{center}}\label{Tab:data1PVllTbsWE}
%\end{table}
%\begin{table}[htb]
\caption{Branching ratios for the $B\to PV\ell^+\ell^-$ decays with the  finite width effects  under the $b\to d \ell^+\ell^-$ transitions within  1$\sigma$ error. Ones for $\ell=e,\mu$ are  in unit of $10^{-7}$, and ones for $\ell=\tau$ are in unit of $10^{-9}$. $^\sharp$denotes that the results should be regards as qualitative estimates since we used the  fixed total width.}
\centering{\footnotesize
\begin{center}  \renewcommand\arraystretch{1}\tabcolsep0.25in
\begin{tabular}{lccc}\hline\hline
~~~~~~~~~~  Decay modes   &       $\mathcal{B}$ with  $\ell=e$        &      $\mathcal{B}$ with   $\ell=\mu$     &    $\mathcal{B}$  with $\ell=\tau$     \\ \hline
$B^+_u\to a^{+}_{2}(\to \rho^{0}\pi^+)\ell^+\ell^-$                &     $0.051\pm 0.018$                  &     $0.034\pm 0.006$                  &     $0.034\pm 0.014$  \\
$B^+_u\to a^{+}_{2}(\to \rho^{+}\pi^0)\ell^+\ell^-$                &     $0.052\pm 0.018$                  &     $0.035\pm 0.007$                  &     $0.034\pm 0.015$  \\
$B^+_u\to a^{+}_{2}(\to \overline{K}^{*0}K^+)\ell^+\ell^-$         &     $(1.18\pm 0.53)\times10^{-4}$$^\sharp$      &     $(7.31\pm 2.14)\times10^{-5}$$^\sharp$      &     $(1.97\pm 0.78)\times10^{-5}$$^\sharp$   \\
$B^+_u\to a^{+}_{2}(\to K^{*+}\overline{K}^0)\ell^+\ell^-$         &     $(1.24\pm 0.54)\times10^{-4}$$^\sharp$      &     $(7.90\pm 2.21)\times10^{-5}$$^\sharp$      &     $(2.18\pm 0.88)\times10^{-5}$$^\sharp$   \\  \hline
$B^0_d\to a^{0}_{2}(\to \rho^-\pi^+)\ell^+\ell^-$                  &     $0.024\pm 0.008$                  &     $0.016\pm 0.003$                  &     $0.016\pm 0.007$  \\
$B^0_d\to a^{0}_{2}(\to \rho^+\pi^-)\ell^+\ell^-$                  &     $0.024\pm 0.008$                  &     $0.016\pm 0.003$                  &     $0.016\pm 0.007$  \\
$B^0_d\to a^{0}_{2}(\to K^{*+}K^-/K^{*-}K^+)\ell^+\ell^-$          &     $(3.15\pm 1.36)\times10^{-5}$$^\sharp$      &     $(2.00\pm 0.56)\times10^{-5}$$^\sharp$      &     $(5.67\pm 2.30)\times10^{-6}$$^\sharp$   \\
$B^0_d\to a^{0}_{2}(\to K^{*0}\overline{K}^0/\overline{K}^{*0}K^0)\ell^+\ell^-$        &     $(2.49\pm 1.12)\times10^{-5}$$^\sharp$      &     $(1.54\pm 0.46)\times10^{-5}$$^\sharp$     &     $(4.07\pm 1.62)\times10^{-6}$$^\sharp$  \\\hline
$B^0_d\to f_{2}(\to K^{*+}K^-/K^{*-}K^+)\ell^+\ell^-$              &   $(8.35\pm 4.12)\times10^{-7}$$^\sharp$       &   $(4.72\pm 1.56)\times10^{-7}$$^\sharp$        &   $(5.29\pm 2.20)\times10^{-8}$$^\sharp$               \\
$B^0_d\to f_{2}(\to K^{*0}\overline{K}^0/\overline{K}^{*0}K^0)\ell^+\ell^-$            &   $(7.35\pm 3.67)\times10^{-7}$$^\sharp$        &   $(4.12\pm 1.37)\times10^{-7}$$^\sharp$        &   $(4.18\pm 1.75)\times10^{-8}$$^\sharp$               \\
$B^0_d\to f_{2}(\to K^{*}K)\ell^+\ell^-$                          &   $(3.14\pm 1.56)\times10^{-6}$$^\sharp$        &   $(1.77\pm 0.59)\times10^{-6}$$^\sharp$        &   $(1.90\pm 0.78)\times10^{-7}$$^\sharp$                \\  \hline
$B^0_d\to f'_{2}(\to K^{*+}K^-/K^{*-}K^+)\ell^+\ell^-$             &   $(2.39\pm 1.01)\times10^{-3}$$^\sharp$        &   $(1.49\pm 0.44)\times10^{-3}$$^\sharp$        &   $(1.50\pm 0.63)\times10^{-4}$$^\sharp$              \\
$B^0_d\to f'_{2}(\to K^{*0}\overline{K}^0/\overline{K}^{*0}K^0)\ell^+\ell^-$           &   $(2.11\pm 0.89)\times10^{-3}$$^\sharp$        &   $(1.31\pm 0.39)\times10^{-3}$$^\sharp$       &   $(1.27\pm 0.53)\times10^{-4}$$^\sharp$             \\
$B^0_d\to f'_{2}(\to K^{*}K)\ell^+\ell^-$                         &   $(9.00\pm 3.81)\times10^{-3}$$^\sharp$       &   $(5.61\pm 1.66)\times10^{-3}$$^\sharp$       &   $(5.54\pm 2.32)\times10^{-4}$$^\sharp$                           \\  \hline
$B^0_s\to \overline{K}^{*0}_{2}(\to K^{*-}\pi^+)\ell^+\ell^-$              &     $0.020\pm 0.007$                  &     $0.013\pm 0.003$                  &     $0.011\pm 0.005$  \\
$B^0_s\to \overline{K}^{*0}_{2}(\to \overline{K}^{*0}\pi^0)\ell^+\ell^-$   &     $(9.68\pm 3.50)\times10^{-3}$                  &     $(6.37\pm 1.27)\times10^{-3}$     &     $(5.34\pm 2.38)\times10^{-3}$  \\
$B^0_s\to \overline{K}^{*0}_{2}(\to \rho^+K^-)\ell^+\ell^-$                &     $(6.96\pm 2.65)\times10^{-3}$     &     $(4.50\pm 1.03)\times10^{-3}$     &     $(3.36\pm 1.47)\times10^{-3}$  \\
$B^0_s\to \overline{K}^{*0}_{2}(\to \rho^0\overline{K}^0)\ell^+\ell^-$     &     $(3.36\pm 1.28)\times10^{-3}$     &     $(2.17\pm 0.50)\times10^{-3}$     &     $(1.60\pm 0.70)\times10^{-3}$  \\
$B^0_s\to \overline{K}^{*0}_{2}(\to \omega\overline{K}^0)\ell^+\ell^-$     &     $(3.30\pm 1.63)\times10^{-3}$     &     $(2.11\pm 0.83)\times10^{-3}$     &     $(1.31\pm 0.69)\times10^{-3}$  \\
$B^0_s\to \overline{K}^{*0}_{2}(\to \overline{K}^{*0}\eta)\ell^+\ell^-$    &     $(5.52\pm 2.22)\times10^{-4}$$^\sharp$      &     $(3.40\pm 0.85)\times10^{-4}$$^\sharp$      &     $(1.21\pm 0.54)\times10^{-4}$$^\sharp$   \\
$B^0_s\to \overline{K}^{*0}_{2}(\to \phi \overline{K}^{0})\ell^+\ell^-$    &     $(3.58\pm 1.65)\times10^{-5}$$^\sharp$      &     $(2.18\pm 0.59)\times10^{-5}$$^\sharp$      &     $(3.63\pm 1.46)\times10^{-6}$$^\sharp$   \\ \hline 		
\end{tabular}
\end{center}}\label{Tab:data1PVllTbdWE}
\end{table}

The results including the finite width effects for the $B\to T(\to PP)\ell^+\ell^-$ and $B\to T(\to PV)\ell^+\ell^-$ decays  are listed in Tabs. \ref{Tab:data1PPllTbsWE}-\ref{Tab:data1PPllTbdWE}  and Tabs. \ref{Tab:data1PVllTbsWE}-\ref{Tab:data1PVllTbdWE}, respectively. Compared with the narrow width results, the numerical results show several characteristic finite width effects in the $B\to T(\to PP)\ell^+\ell^-$  and $B\to T(\to PV)\ell^+\ell^-$ decays.
\begin{enumerate}
\item
The finite width effects slightly reduce the branching ratios for most $PP$ and $PV$ channels because their thresholds lie well below the tensor resonance masses  and their  phase spaces vary smoothly across the resonance regions. Therefore, the finite width integration mainly smears the resonance contribution over the allowed invariant-mass range, resulting in slightly smaller branching fractions in most cases.

\item  Decay $f_2'(1525)\to\eta\eta'$  arises from nearby thresholds and the $D$-wave phase-space dependence ($\Gamma\propto p^5$). The $\eta\eta'$ threshold lies only about $11.7~\mathrm{MeV}$ below the average $f_2'(1525)$ mass, so the finite width integration includes the high-mass region where the decay momentum increases rapidly. Within the present fixed-width, the finite width integration gives a larger branching fraction for the $f_2'(1525)\to\eta\eta'$ related  channels such as $B_s^0\to f_2'(1525)(\to\eta\eta')\ell^+\ell^-$ and $B_d^0\to f_2'(1525)(\to\eta\eta')\ell^+\ell^-$ decays. A detailed discussion can be found in the analysis of $K_2^*(1430)\to K\eta'$ given below.

\item
For $f_2(1270)\to\eta\eta$ and $a_2(1320)\to\pi\eta'$ decays, their threshold energy differences are about 200 MeV,
these decays are farther from threshold than $f_2'(1525)\to\eta\eta'$, and their finite width corrections are
therefore less pronounced.  For the $e^+e^-$ and $\mu^+\mu^-$ modes, the branching fraction can be slightly increased or unchanged by the $p^{5}$ factor.
In contrast, the $\tau^+\tau^-$ modes are suppressed because the lower limit
\(q^2_{\min}=4m_\tau^2\) is already close to the kinematic upper limit    $q^2_{\max}(t) = \left(m_B-\sqrt{t}\right)^2$.
An increase of the off-shell tensor mass therefore strongly reduces
the available dilepton phase space.

\item
For $B\to T(\to PV)\ell^+\ell^-$  processes, since we have considered the finite width effects of both vector and tensor mesons in the final-state, the width correction effects suppress the branching ratios of all processes, and the suppression is quite significant in some cases.

\item In addition,  for the subthreshold $K_2^*(1430)\to K\eta'$ decay, the $K\eta'$ threshold lies about $24~\mathrm{MeV}$ above the  $K_2^*$ mass, and hence the decays are forbidden in the narrow width approximation. The high-mass tail of the broad $K_2^*$ resonance nevertheless extends into the physical $K\eta'$ region and produces   nonzero finite width contribution.  In this case, $\mathcal{B}(K_2^*\to K\eta',t_R)=\Gamma(K_2^*\to K\eta',t_R)/\Gamma_{K_2^*}$ in Eq. (\ref{Eq:RDA4})  may become unphysically larger than the corresponding inclusive tensor resonance contributions if the rapidly increasing $K\eta'$ partial width is combined with a fixed total
width. We therefore do not give the corresponding numerical results.  A consistent calculation requires an energy-dependent total width satisfying
$0\leq\frac{\Gamma(K_2^*\to K\eta',t_R)}{\Gamma_{K_2^*}^{\rm tot}(t_R)}\leq 1.$   A reliable construction of the energy-dependent total width requiresadditional information on all relevant decay channels and is beyond
the scope of the present analysis.
The use of a fixed total width is expected to have its largest impacton near-threshold and subthreshold channels. Therefore, the numericalresults for such channels should be regarded as qualitative estimates, which are denoted by $^\sharp$ in Tabs. \ref{Tab:data1PPllTbsWE}-\ref{Tab:data1PVllTbdWE} .
A quantitatively consistent treatment requires an energy-dependent total width constructed from all relevant open decay channels.

\end{enumerate}
In brief summary, the finite width corrections result from the competition between the increase of the hadronic $D$-wave phase space, proportional to $p^5(t)$, and the reduction of the available $B\to T\ell^+\ell^-$ phase space as the off-shell tensor mass increases. The former effect is generally more important for the $e^+e^-$ and $\mu^+\mu^-$ modes close to a hadronic threshold, while the latter can also be significant  in the $\tau^+\tau^-$ modes.

Finally, there are two additional points that need to be noted.
\begin{itemize}
\item The results  in Tabs. \ref{Tab:data1PPllTbs}-\ref{Tab:data1PVllTbdWE} represent individual tensor resonance contributions. Interference effects may become relevant when two or more resonance amplitudes contribute to the same final state with comparable magnitudes and overlapping invariant-mass distributions. Their quantitative size also depends on the relative strong phases.  Consequently, the decay rate contains both the individual resonance contributions and their interference terms.
 Such interference effects depend on the resonance line shapes and relative strong phases, since these strong phases are not constrained by the currently available data, a reliable quantitative evaluation of the interference effects is not possible within the present framework.
Nevertheless,  Tab.\ref{tab:samefinalPPPV} 	lists the final states  receiving contributions from more than one tensor resonance and indicates the dominant tensor component inferred	from our numerical results.

\item  For the normalization channel $B_s^0\to f_2'(1525)\mu^+\mu^-$ decay, possible long-distance contributions to this decay are not separately determined by the available measurement and are therefore not propagated independently in our numerical analysis. They should be regarded as an additional source of systematic uncertainty in the extracted normalization and in the resulting branching-fraction estimates.

\end{itemize}

\begin{table}[htbp]
\centering
\caption{Tensor resonances contributing to the same $B\to PP\ell^+\ell^-$ and $B\to PV\ell^+\ell^-$  final states and the corresponding	dominant contributions. }	 \label{tab:samefinalPPPV}
\small
\setlength{\tabcolsep}{5pt}
\renewcommand{\arraystretch}{1.1}
\begin{tabular}{		p{0.29\textwidth}		p{0.29\textwidth}   		 p{0.34\textwidth}    	}
\toprule \hline \hline
Decay modes 	&	Possible tensor resonances	&	Dominant contribution	 \\\hline
\midrule	
$B_s^0\to\pi^+\pi^-\ell^+\ell^-,\pi^0\pi^0\ell^+\ell^-$	&	$f_2(1270), f_2'(1525)$	&	Dominated by $f_2(1270)$	\\
$B_s^0\to K^+K^-\ell^+\ell^-,K^0\bar K^0\ell^+\ell^-,$     	&	 $f_2(1270)$, $f_2'(1525)$	&	Strongly dominated by $f_2'(1525)$	\\
 $B_s^0\to\eta\eta\ell^+\ell^-,~B_d^0\to\eta\eta\ell^+\ell^-$    	&	 $f_2(1270)$, $f_2'(1525)$	&	Strongly dominated by $f_2'(1525)$	\\
$B_d^0\to\pi^+\pi^-\ell^+\ell^-,\pi^0\pi^0\ell^+\ell^-$	&	$f_2(1270)$, $f_2'(1525)$	&	Dominated by $f_2(1270)$	\\
$B_d^0\to K^+K^-\ell^+\ell^-,K^0\bar K^0\ell^+\ell^-$   	&	 $a_2^0(1320)$, $f_2(1270)$, $f_2'(1525)$	&	$f_2'(1525)$ dominates the $e$ and $\mu$ modes,	 whereas the $f_2'(1525)$ and $a_2^0(1320)$	 contributions are comparable in the $\tau$ mode\\	\hline
$B_s^0\to K^{*+}K^-\ell^+\ell^-,K^{*-}K^+\ell^+\ell^-$			&			 $f_2(1270)$,$f_2'(1525)$			&			Strongly dominated by $f_2'(1525)$			 \\
$B_s^0\to K^{*0}\bar K^0\ell^+\ell^-$, $\bar K^{*0}K^0\ell^+\ell^-$	&			 $f_2(1270)$,$f_2'(1525)$			&			Strongly dominated by $f_2'(1525)$			\\
$B_d^0\to K^{*+}K^-\ell^+\ell^-,K^{*-}K^+\ell^+\ell^-		$	&			 $a_2^0(1320)$,$f_2(1270)$,$f_2'(1525)$	&		Dominated by $f_2'(1525)$			 \\
$B_d^0\to K^{*0}\bar K^0\ell^+\ell^-,\bar K^{*0}K^0\ell^+\ell^-$	&			 $a_2^0(1320)$,$f_2(1270)$,$f_2'(1525)$	&		Dominated by $f_2'(1525)$			 \\\hline
\end{tabular}
\end{table}

\newpage
\section{Summary}

The tensor resonance contributions in the $B\to PP\ell^+\ell^-,PV\ell^+\ell^-$ (with $\ell=e,\mu,\tau)$ decays with  $b\to s/d\ell^+\ell^-$ quark level transitions  have been studied  based on the SU(3) flavor symmetry approach.
The  branching ratios of  the $B\to T(\to PP)\ell^+\ell^-$ and $B\to T(\to PV)\ell^+\ell^-$ decays  have been evaluated  both in the narrow width approximation and with finite width corrections.

The hadronic amplitudes of the $B \to T\ell^+\ell^-$ decays are related by the SU(3) flavor symmetry. Using the only experimental data of  $\mathcal{B}(B^0_s\to f^{\prime}_{2}(1525)\mu^+\mu^-)$,  all branching ratios of the $B \to T\ell^+\ell^-$ decays are obtained  with the assistance of the $B\to T$ form factors  from PQCD and LCSR.
We found that phenomenological estimates of the branching ratios  of  all $B\to T\tau^+\tau^-$ decays in $S_{LCSR}$ case are slightly smaller than those in $S_1$ and $S_{PQCD}$ cases. Other branching ratio estimates of the  $B \to T\ell^+\ell^-$ decays are consistent with each other within $1\sigma$ error bar. $\mathcal{B}(B_s^0\to f_2(1270)\ell^+\ell^-)$ and $\mathcal{B}(B_d^0\to f'_2(1525)\ell^+\ell^-)$ are given for the first time to our knowledge.
In addition, the branching ratios, the longitudinal polarization fractions, the normalized forward-backward asymmetries,  and some other observables   with $\ell=e,\mu$ and the different $q^2$ bins are obtained,
and  the observables of the $B\to T\tau^+\tau^-$ decays  are also obtained in the whole $q^2$ ranges.

For the branching ratios of  $B\to T(T\to PP)\ell^+\ell^-$, we found that, in the decays with $\ell=e,\mu$, only $\mathcal{B}(B^0_s\to f'_2(\to K^0\overline{K}^0)e^+e^-)$ and $\mathcal{B}(B^0_s\to f'_2(\to K^+K^-)e^+e^-)$ are on the order of $10^{-7}$, the others are on the order of $10^{-8}$ or smaller.
Our prediction  $\mathcal{B}(B^0_s\to f'_2(\to \pi^+\pi^-)\mu^+\mu^-)$  is much smaller than corresponding total experimental  $\mathcal{B}(B^0_s\to  \pi^+\pi^-\mu^+\mu^-)$.  Similarly, our prediction of $\mathcal{B}(B^0_d\to f_2(\to \pi^+\pi^-)\mu^+\mu^-)$ and $\mathcal{B}(B^0_d\to f'_2(\to \pi^+\pi^-)\mu^+\mu^-)$  are also  much smaller than  corresponding total experimental  $\mathcal{B}(B^0_d\to  \pi^+\pi^-\mu^+\mu^-)$. These may mean that the scalar meson  or vector meson   resonances rather than the tensor resonances give the dominant contributions in some $B\to PP \ell^+\ell^-$ decays.
In addition,  approximately half of the considered modes have branching ratios of  order of $10^{-10}$ in the $B\to T(\to PP)\tau^+\tau^-$  decays, and other ones are on the order of $\mathcal{O}(10^{-14})-\mathcal{O}(10^{-11})$.

For the $B\to T(\to PV)\ell^+\ell^-$ decays, their branching ratio estimates  have been predicted  on the order of $\mathcal{O}(10^{-13})-\mathcal{O}(10^{-8})$ for $\ell=e,\mu$ and $\mathcal{O}(10^{-16})-\mathcal{O}(10^{-10})$ for $\ell=\tau$. The current experimental measurement $\mathcal{B}(B^+_u\to  \phi K^+\mu^+\mu^-)$ is much larger than our prediction $\mathcal{B}(B^+_u\to K^{*+}_2(\to \phi K^+)\mu^+\mu^-)$. This indicates that other resonances such as axial-vector mesons or their excited states may give the dominant contributions to some  $B \to  PV\ell^+\ell^-$ decays.

 Compared with the narrow width results,   the finite width effects slightly reduce the branching fractions for most $PP$ and $PV$ channels.
However, sizeable finite width effects are found in the $f_2'(1525)\to\eta\eta'$ channels.
The finite width integration gives a larger branching fraction
than the narrow width approximation for $B_s^0\to f_2'(1525)(\to\eta\eta')\ell^+\ell^-$ and $B_d^0\to f_2'(1525)(\to\eta\eta')\ell^+\ell^-$ decays with $\ell=e,\mu$. The high-mass tail of the broad $K_2^{*}(1430)$ gives
nonzero finite width contributions in $B_u^+\to K_2^{*+}(1430)(\to K^+\eta')\ell^+\ell^-$ and $B_d^0\to K_2^{*0}(1430)(\to K^0\eta')\ell^+\ell^-$ decays,  however, a reliable quantitative prediction requires an energy-dependent total width, and therefore no numerical result  is quoted for these channels.

Based on our phenomenological estimates, some of the modes with relatively large branching ratios may be useful targets for future experimental searches, and
our results can be tested in current and future experiments.

\section*{ACKNOWLEDGEMENTS}
The work was supported by the National Natural Science Foundation of China (No. 12365014).

\section*{References}

\end{document}